\documentclass[twocolumn]{aastex631}
\usepackage{comment}
\usepackage{natbib} 
\usepackage{amsmath}
\usepackage[flushleft]{threeparttable}

\newcommand{\kms}{km s$^{-1}$}

\newcommand{\hii}{H~{\scriptsize II}}

\newcommand{\msun}{M$_{\odot}$}

\newcommand{\degree}{$^{\circ}$}

\graphicspath{{./}{figures/}}

\defcitealias{Butterfield2024a}{FIREPLACE I}
\defcitealias{Butterfield2024b}{FIREPLACE II}
\defcitealias{Pare2024}{FIREPLACE III}

\begin{document}

\title[FIREPLACE IV (DR2)]{\uppercase{SOFIA/HAWC+ Far-Infrared Polarimetric Large-Area CMZ Exploration Survey. IV. Relative Magnetic Field Orientation Throughout the CMZ}}

\correspondingauthor{Dylan M. Par\'e}
\email{dylanpare@gmail.com}

\author[0000-0002-5811-0136]{Dylan M. Par\'e}
\affiliation{Department of Physics, Villanova University, 800 E. Lancaster Ave., Villanova, PA 19085, USA}

\author[0000-0003-0016-0533]{David T. Chuss}
\affiliation{Department of Physics, Villanova University, 800 E. Lancaster Ave., Villanova, PA 19085, USA}

\author[0009-0006-4830-163X]{Kaitlyn Karpovich}
\affiliation{Department of Physics, Stanford University, Stanford, CA 94305, USA}
\affiliation{Kavli Institute for Particle Astrophysics \& Cosmology, P.O. Box 2450, Stanford University, Stanford, CA 94305, USA}
\affiliation{Department of Physics, Villanova University, 800 E. Lancaster Ave., Villanova, PA 19085, USA}

\author[0000-0002-4013-6469]{Natalie O. Butterfield}
\affiliation{National Radio Astronomy Observatory, 520 Edgemont Road, Charlottesville, VA 22903, USA}

\author[0000-0001-7466-0317]{Jeffrey Inara Iuliano}
\affiliation{Department of Physics, Villanova University, 800 E. Lancaster Ave., Villanova, PA 19085, USA}

\author[0000-0003-1337-9059]{Xing Pan}
\affiliation{School of Astronomy and Space Science, Nanjing University, 163 Xianlin Avenue, Nanjing 210023, People's Republic of China}
\affiliation{Key Laboratory of Modern Astronomy and Astrophysics (Nanjing University), Ministry of Education, Nanjing 210023, People's Republic of China}
\affiliation{Center for Astrophysics $|$ Harvard \& Smithsonian, 60 Garden Street, Cambridge, MA 02138, USA}

\author[0000-0002-7567-4451]{Edward J. Wollack}
\affiliation{NASA Goddard Space Flight Center, Mail Code: 665, Greenbelt, MD 20771}

\author[0000-0003-2384-6589]{Qizhou Zhang}
\affiliation{Center for Astrophysics $|$ Harvard \& Smithsonian, 60 Garden Street, Cambridge, MA 02138, USA}

\author[0000-0002-6753-2066]{Mark R. Morris}
\affiliation{Department of Physics \& Astronomy, University of California, Los Angeles, 475 Portola Pl., Los Angeles, CA 90095-1547, USA}

\author[0009-0008-8494-8777]{Mathilda Nilsson}
\affiliation{Department of Physics and Astronomy, University of Pittsburgh, 100 Allen Hall, 3941 O'Hara St., Pittsburgh, PA 15260}

\author[0009-0001-9716-4188]{Roy J. Zhao}
\affiliation{Kavli Institute for Cosmological Physics, The University of Chicago, 5640 S Ellis Ave., Chicago, IL 60637, USA}
\affiliation{Department of Physics, The University of Chicago, 5720 S Ellis Ave., Chicago, IL 60637, USA}
\affiliation{Department of Physics \& Astronomy, University of California, Los Angeles, 475 Portola Pl., Los Angeles, CA 90095-1547, USA}

\begin{abstract}

The nature of the magnetic field structure throughout the Galactic Center (GC) has long been of interest. The recent Far-InfraREd Polarimetric Large-Area CMZ Exploration (FIREPLACE) Survey reveals preliminary connections between the seemingly distinct vertical and horizontal magnetic field distributions previously observed in the GC. We use the statistical techniques of the Histogram of Relative Orientation (HRO) and the Projected Rayleigh Statistic (PRS) to assess whether the CMZ magnetic field preferentially aligns with the structure of the CMZ molecular clouds or the morphology of the non-thermal emission of the GC NTF population. We find that there is a range of magnetic field orientations throughout the population of CMZ molecular clouds, ranging from parallel to perpendicular orientation. We posit these orientations depend on the prevalence of gravitational shear in the GC in contrast with what is observed in Galactic Disk star-forming regions. We also compare the magnetic field orientation from dust polarimetry with individual prominent NTFs, finding a preferred perpendicular relative orientation. This perpendicular orientation indicates that the vertical field component found in the FIREPLACE observations is not spatially confined to the NTFs, providing evidence for a more pervasive vertical field in the GC. From dynamical arguments, we estimate an upper limit on the magnetic field strength for this vertical field, finding $B \leq 4$ mG. A field close to this upper limit would indicate that the NTFs are not local enhancements of a weaker background field and that the locations of the NTFs depend on proximity to sites of cosmic ray production.
 
\end{abstract}


\section{INTRODUCTION} \label{sec:intro}
\begin{figure*}
    \centering
    \includegraphics[width=1.0\textwidth]{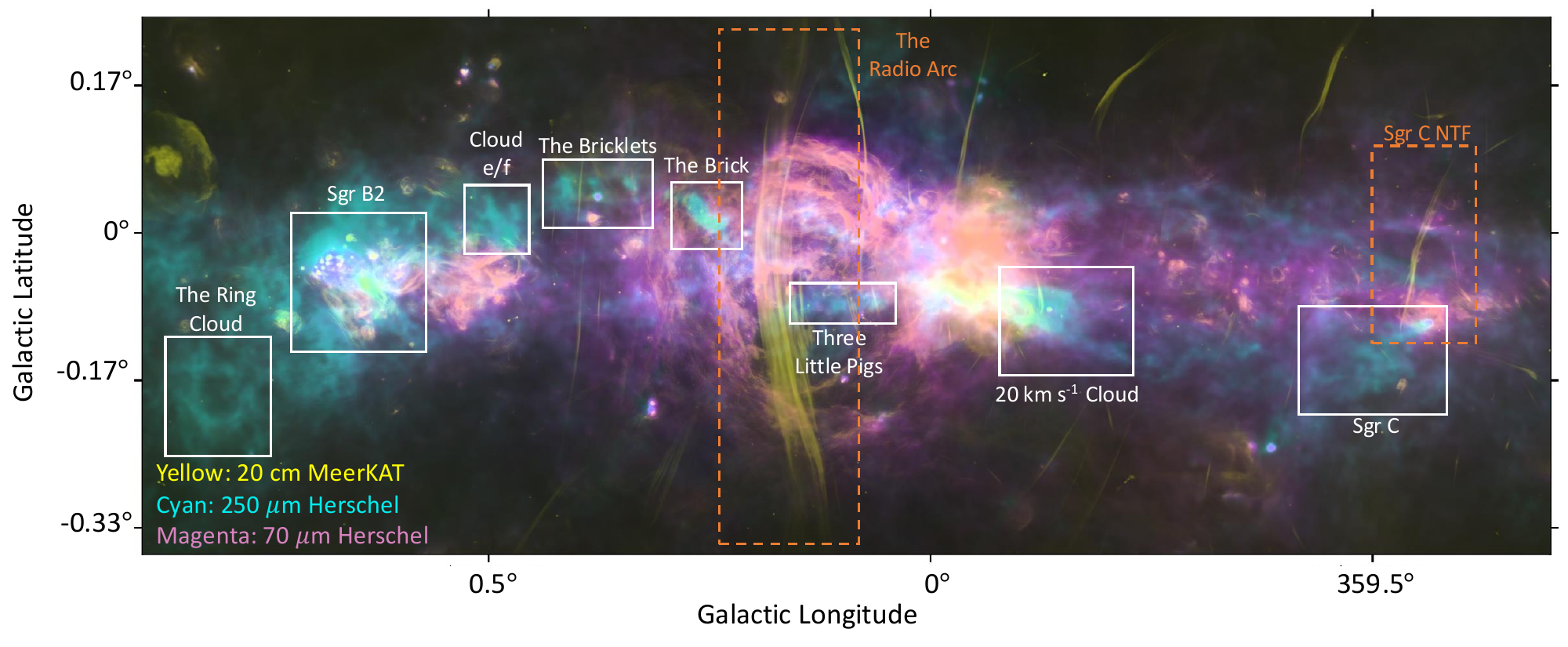}
    \caption{The CMZ as illustrated by radio and infrared observations. This 3-color image shows the 20 cm (1.28 GHz) MeerKAT observations revealing the hot magnetized plasma \citep[yellow,][]{Heywood2022} and the 70 \micron\ and 250 \micron\ Herschel observations tracing warm and cool dust in the CMZ \citep[magenta and cyan, respectively,][]{Molinari2010a}. Prominent molecular clouds and non-thermal structures studied in this work are labeled and marked with white and orange boxes, respectively.}
    \label{fig:legend}
\end{figure*}

The Galactic Center (GC) is a unique region of the Milky Way that possesses a large amount of dense gas \citep[2--6$\times$10$\rm^7$ \msun,][]{Morris1996b,Barnes2017}. This dense gas is ordered into a population of molecular clouds within the central 150 pc that are collectively known as the Central Molecular Zone (CMZ). A 3-color view of the CMZ, as probed by infrared and radio observations, is shown in Figure \ref{fig:legend}.

Despite the high densities in the CMZ, the star formation efficiency (SFE) is only 10\% that of the Galactic Disk \citep{Longmore2013,Morris2023}. The exact reason for this depressed SFE is unknown, but multiple possibilities have been posited such as the elevated turbulence \citep{Krumholz2015} and high magnetic field strengths \citep[100s of $\mu$G -- 10s of mG, e.g.][]{Pillai2015,Mangilli2019,Lu2024} of the region. Alternatively, the CMZ could be in a period of relative inactivity \citep{Kruijssen2014}. Of these mechanisms, the role of the magnetic field has remained the most uncertain.

There are multiple ways to probe the nature of the magnetic field in the GC. At radio wavelengths, a population of linear, synchrotron structures is observed that are known as the non-thermal filaments (NTFs). Multiple NTFs can be observed in Figure \ref{fig:legend}, where the NTFs appear as long coherent structures in the yellow synchrotron emission. Two prominent NTFs are marked with orange boxes in Figure \ref{fig:legend}. The magnetic fields inferred for individual NTFs are generally oriented parallel to the NTF filaments \citep[e.g.,][]{YWP1997,Lang1999a,Lang1999b}. This finding, coupled with the fact that the NTF population is generally oriented perpendicular to the Galactic plane \citep{Yusef-Zadeh2022}, implies the presence of a vertical magnetic field across the GC. However, observations at other wavelength regimes paint a different picture.

The magnetic field local to dust grains can be inferred from polarimetric observations at sub-millimeter and far-infrared wavelengths \citep[e.g.,][]{Hildebrand2009}. At arcminute resolutions the field inferred from dust polarization is fairly uniform and could be a foreground field in the central kpc of the Galaxy \citep{Mangilli2019,Guan2021}. At higher resolutions (19\farcs6), the field observed at far-infrared wavelengths is highly coupled to CMZ dynamics as observed by the Far-Infrared Polarimetric Large-Area CMZ Exploration (FIREPLACE, PI: D. Chuss) Survey using the Stratospheric Observatory for Infrared Astronomy \citep[SOFIA,][herafter referred to as FIREPLACE I and FIREPLACE III, respectively]{Butterfield2024a,Pare2024}.

At 214 $\mu$m the inferred polarization is largely likely due to the Radiative Torque Alignment mechanism \citep{Lazarian07,Andersson2015}. The 90 GHz observations of \citet{Guan2021} observe potential mixing between dust polarization and synchrotron emission. However, at 214 $\mu$m, there is likely not a significant contribution from synchrotron emission due to its steep spectrum.

In this work, we follow up on the preliminary magnetic field alignment analysis presented in \citetalias{Pare2024} with a quantitative analysis of how the magnetic field inferred from the SOFIA/HAWC+ FIREPLACE observations aligns with the orientations of the CMZ molecular clouds and the GC NTFs. We utilize the Histogram of Relative Orientation \citep[HRO,][]{Soler2013} and Projected Rayleigh Statistic \citep[PRS,][]{Jow2018} algorithms to statistically evaluate the alignment of the FIREPLACE magnetic field with the CMZ molecular cloud column density distribution derived from Herschel and the Radio NTF emission from MeerKAT.

In Section \ref{sec:obs} we detail the observations used to conduct our analysis. In Section \ref{sec:tech} we briefly summarize the HRO and PRS algorithms used to quantitatively assess the magnetic field alignment. Section \ref{sec:cmz_align} presents the results on the alignment of the FIREPLACE magnetic field with the dust column density and radio synchrotron emission throughout the entire CMZ. Section \ref{sec:cloud_align} builds on the CMZ-wide results presented in Section \ref{sec:cmz_align} by analyzing the relative alignment between the magnetic field and the column density structure of prominent CMZ molecular clouds and individual NTFs. In Section \ref{sec:disc} the implications of these results are discussed and the key findings of this study are summarized in Section \ref{sec:conc}.

\section{OBSERVATIONS} \label{sec:obs}
\begin{figure*}
    \centering
    \includegraphics[width=1.0\textwidth]{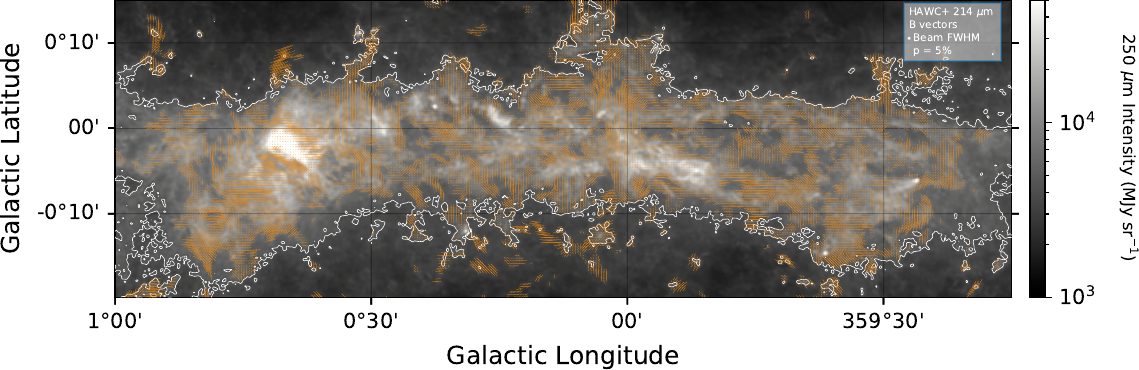}
    \caption{The magnetic field derived from infrared dust polarization for the CMZ. The FIREPLACE magnetic field pseudovectors are shown as orange line segments. The greyscale background is 250 \micron\ Herschel emission of the CMZ \citep{Molinari2011}.}
    \label{fig:fire_B}
\end{figure*}
We use three data products to conduct the analysis presented in this paper: magnetic field direction information derived from the far-infrared polarimetric FIREPLACE observations of the CMZ, the column density distribution of the molecular clouds in the CMZ derived from multi-wavelength Herschel observations, and the intensity distribution of the population of NTFs throughout the GC from MeerKAT 1.28 GHz observations of the region. The combination of these data products enables a quantitative study of the alignment between the magnetic field orientation with the thermal and non-thermal structures in the GC.

\subsection{FIREPLACE 214 \micron\ Observations}
The dust polarization magnetic field data are obtained from the second data release (DR2) of the FIREPLACE survey of the CMZ. The reduction methods and initial presentation of the FIREPLACE DR2 data set is presented in \citetalias[][]{Pare2024}. We follow the significance cuts employed in that paper where pseudovectors are Nyquist-sampled and are only determined for lines-of-sight that satisfy the total and polarimetric cuts of $I/\sigma_I >$ 200, $p_{\%} < 50$\%, and $p/\sigma_p >$ 3 where $I$ and $\sigma_I$ are the FIREPLACE 214 \micron\ intensity and uncertainty, $p$ and $\sigma_p$ are the FIREPLACE 214 \micron\ debiased polarized intensity and uncertainty, and $p_{\%}$ is the FIREPLACE percentage polarization. These significance cuts are the standard SOFIA polarimetry cuts \citep{Gordon2018}. The number of Nyquist-sampled polarization measurements obtained for the CMZ after these cuts is $\sim$64,000. The distribution of the FIREPLACE magnetic field pseudovectors throughout the CMZ is shown in Figure \ref{fig:fire_B}.

\subsection{Herschel Derived N(H$_2$) Column Density}
\begin{figure*}
    \centering
    \includegraphics[width=1.0\textwidth]{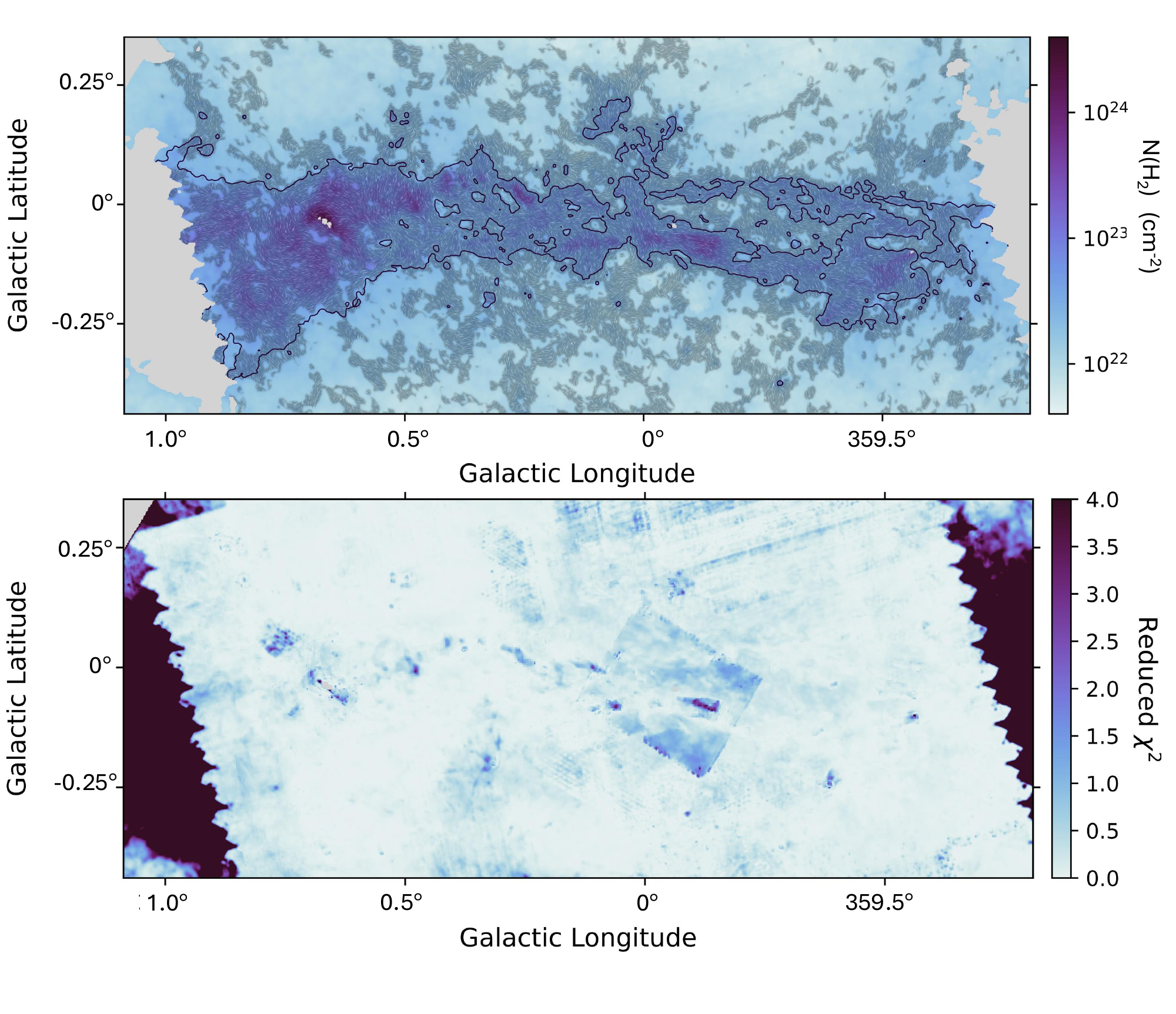}
    \caption{\textit{Top:} Column density distribution corresponding to the CMZ molecular emission. This N(H$_2$) distribution was derived by fitting a modified blackbody curve to the 70, 160, 250, and 350 \micron\ Herschel Hi-Gal observations \citep{Molinari2010a}, as described in the text. Column density values corresponding to lines of sight with $\chi^2$ values $\geq$5 have been masked. The contour level marks $3\times10^{22}$, which is the minimum column density evaluated for the HRO and PRS analysis later in this work. \textit{Bottom:} The $\chi^2$ map used to threshold the column density map shown in the upper panel.}
    \label{fig:NH2}
\end{figure*}
The FIREPLACE-derived magnetic field orientations will be compared to the morphology of the molecular clouds and NTFs in the GC. The distribution of molecular material in the GC is obtained by computing the N(H$_2$) column density. The column density is found by fitting a modified blackbody curve to Herschel observations from the Hi-Gal Survey of the Galactic plane \citep{Molinari2010a}. This fit is performed using the Herschel 70, 160, 250, and 350 \micron\ observations smoothed to the lowest resolution of 25\arcsec\ to match that of the 350 \micron\ data. We followed the method for modified blackbody curve fitting described in \citet{Chuss2019}. Because of the highly uncertain extinction at 70 \micron\ we compared fits with and without the 70 \micron\ data. We obtained similar column density values with both approaches, and we therefore use the results that incorporate the 70 \micron\ observations in this paper.

The intensity was modeled as found in \citet{Vaillancourt2002}:
\begin{equation}
    I_{\nu}=(1-e^{-\tau_{\nu}})B_{\nu}(T), \label{eq:black}
\end{equation}
where $B_{\nu}(T)$ is the Planck blackbody function at frequency $\nu$ and temperature $T$. The optical depth, $\tau(\nu)$, is defined as $\tau \equiv \epsilon(\frac{\nu}{\nu_0})^{\beta}$, where $\epsilon$ is a constant  of proportionality related to the column density along the line of sight and $\beta$ is the dust emissivity index. This is the same methodology that has been employed in other star-forming regions in the Galaxy \citep{Battersby2011}.

We used a constant dust emissivity index of $\beta = 2$ as was used in \citet{Bernard2010} and \citet{Molinari2011}. Following \citet{Sadavoy2013}, a value of $1000$ GHz was used for $\nu_0$. $\epsilon$ is defined as
\begin{equation}
    \epsilon = \kappa_{\nu_0}\mu m_H N(H_2), \label{eq:eps}
\end{equation}
where $\kappa_{\nu_0}$ is a reference dust opacity per unit mass at $\nu_0$, $\mu$ is the mean molecular weight per hydrogen atom, $m_H$ is the atomic mass of hydrogen, and $N(H_2)$ is the gas column density. We use $\kappa_{\nu_0}=0.1 \:cm^2 \:g^{-1}$ and $\mu = 2.8$ following \citet{Sadavoy2013}. The modified blackbody curve was fit for the temperature and $\epsilon$, which was then converted to column density using Equation \ref{eq:eps}. The resulting Herschel-derived $N(H_2)$ distribution obtained from this fitting is shown in the upper panel of Figure \ref{fig:NH2}. We also present the reduced $\chi^2$ goodness-of-fit metric for the column density fitting in the bottom panel of Figure \ref{fig:NH2}. We mask out column density fits coinciding with reduced $\chi^2$ values that are $\geq$5.0, as shown in the upper panel of Figure \ref{fig:NH2}. We note that the column density fits associated with the higher column density regimes that are analyzed later in the paper generally have reduced $\chi^2 \leq$2.0. The resulting column density distribution is then sub-sampled to the 19\farcs6 resolution of the FIREPLACE observations for the HRO/PRS analysis employed later in this work.

An alternative method of determining the column density was detailed in \citet{Marsh2015} that allows for the determination of column density at higher resolution. However, a critical limit of this technique is that it is only reliable for optically thin structures. Because of the variation of optical density in the CMZ, we do not use this method for generating our column density distribution.

Previous work deriving the column density distribution in the CMZ has been performed using a variable $\beta$ \citep[compared to the assumed constant $\beta$ employed in this work,][]{Tang2021}. We find similar column density values in our column density map to the distribution presented in \citet{Tang2021}. We therefore use the constant $\beta$ map derived in this work since it covers the full CMZ out to Sgr B2.

\subsection{MeerKAT 1.28 GHz Radio Intensity}
\begin{figure*}
    \centering
    \includegraphics[width=1.0\textwidth]{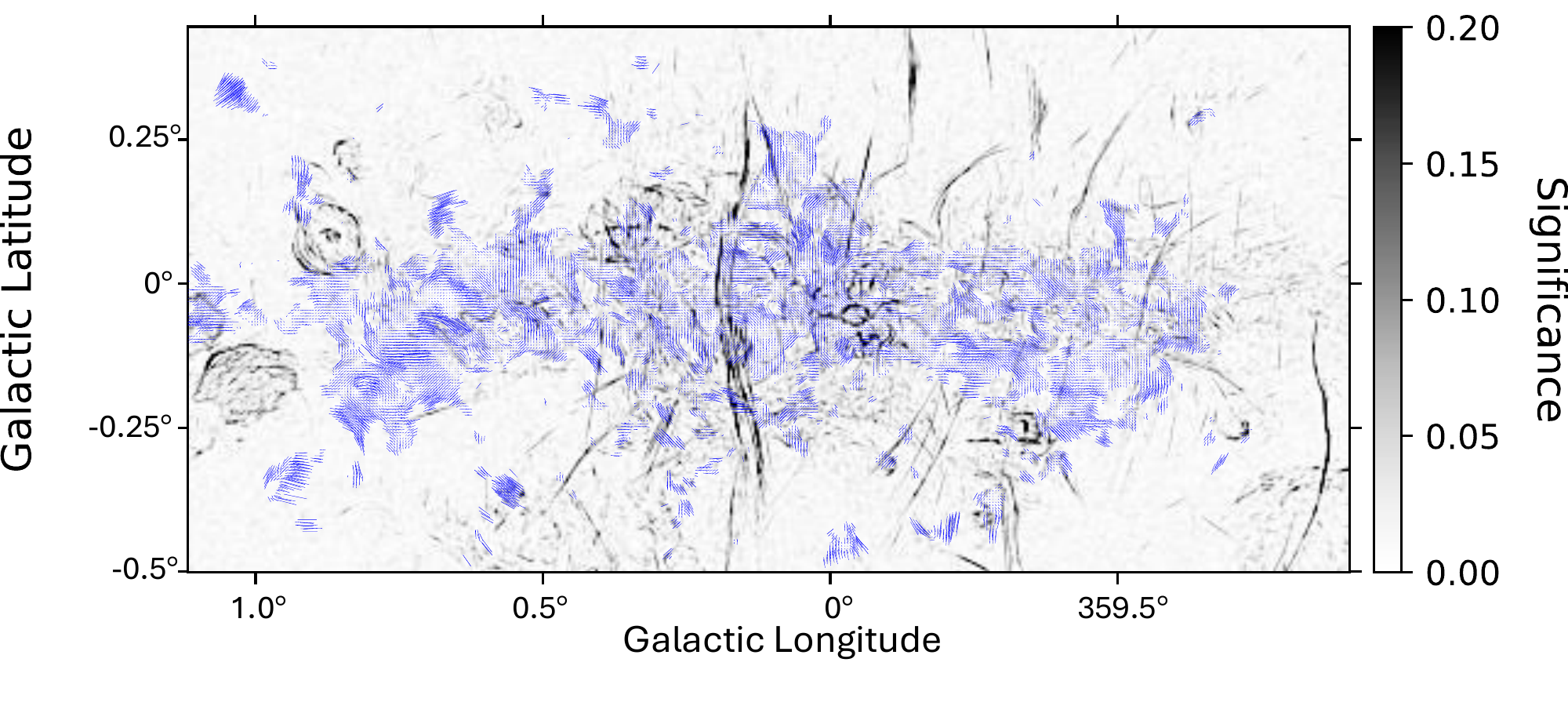}
    \caption{Filtered 1.28 GHz map of the MeerKAT observations of the GC from \citet{Heywood2022}, with FIREPLACE pseudovectors overlayed as blue line segments. The field of view of the radio image is chosen to match the field of view of the column density distribution shown in the upper panel of Figure \ref{fig:NH2}.}
    \label{fig:MeerKAT_NTF}
\end{figure*}
To compare the magnetic field pseudovectors with the NTF distribution in the GC, we use the 1.28 GHz (1\arcsec\ pixel size) MeerKAT observations of the region \citep{Heywood2022}. The MeerKAT observations include emission from a complex combination of structures, but since we are most interested in comparing to the NTF orientation we apply the Rolling Hough Transform (RHT) method introduced and described in \citet{Clark2014} to isolate the emission of the NTF filamentary structures.

The resulting filtered MeerKAT distribution is shown in Figure \ref{fig:MeerKAT_NTF}, and was generated using a smoothing diameter ($D_k$) of 8 pixels ($\sim$8\arcsec\ or 0.3 pc), a window diameter ($D_w$) of 41 pixels ($\sim$41\arcsec\ or 1.6 pc), and a threshold of parameter ($Z$) of 0.7. These parameters are described in detail in \citet{Clark2014}. A range of RHT parameter values were explored to optimize the filtering for the NTFs and the above set of parameters was best able to extract the NTF filaments from the unfiltered MeerKAT distribution. 


\section{TECHNIQUES} \label{sec:tech}
Using the data products described in Section \ref{sec:obs}, we briefly describe our methods for deriving the relative alignment of the magnetic field with the column density and NTF distributions.

\subsection{Deriving Magnetic Field Orientation}
\subsubsection{Histogram of Relative Orientation (HRO)}
The HRO method was first presented in \citet{Soler2013}. We will not detail all of the derivation of the HRO method here, since a more robust discussion can be found in \citet{Soler2013}, but we briefly summarize the key points for the convenience of the reader.

For the HRO, the relative orientation between the polarization angle and the gradient of the column density is determined by comparing the cross and dot products of the two vectors:
\begin{equation}
    \phi = \tan^{-1}\left(\frac{|\mathbf{E}\times\boldsymbol{\nabla} n|}{\mathbf{E}\cdot\boldsymbol{\nabla} n}\right), \label{eq:phi}
\end{equation}
where $\mathbf{E}$ is the polarization, $\boldsymbol{\nabla} n$ is the gradient of the column density, and $\phi$ is the relative angle between the magnetic field orientation and column density gradient. To ensure that we are computing the column density gradient over an appropriate resolution we use a Gaussian kernel ($\sigma$) to smooth the input data set. For the column density gradient we chose a kernel size of $\sigma\sim$14\farcs7 (3 pixels), equivalent to Nyquist sampling the lower resolution column density distribution (with a beam size of 25\farcs0). This is a smaller kernel size than has been used in previous studies \citep[e.g.,][]{Jow2018} where the kernel size is motivated to suppress the noise when determining the gradient. The high signal-to-noise of our column density map and FIREPLACE polarization data \citepalias{Pare2024}, coupled with our focus on higher column density and polarimetric SNR regions, motivates this smaller kernel size.

The distribution of relative angles is then binned to produce a histogram (referred to as the HRO). Significant non-uniform structure in the HRO can therefore be used to determine the extent to which the magnetic field has a preferential orientation relative to the column density gradient. The significance of the HRO peak can be quantified by using the histogram shape parameter:
\begin{equation}
    \zeta = A_c - A_e, \label{eq:zeta}
\end{equation}
where $A_c$ is the area under the curve filled by the central portion of the histogram from the angle range -0.25 $<\cos\phi<$ 0.25, and $A_e$ is the area under the curve filled by the extreme edges of the histogram from the angle range -1.00 $<\cos\phi<$ -0.75 and 0.75 $<\cos\phi<$ 1.00.

The histogram shape parameter can therefore be used to characterize the preferential relative orientation of the magnetic field compared to the column density gradient. If $\zeta>$ 0.0, the peak of the histogram occurs around $\cos\phi\sim$ 0.0, indicating a parallel alignment of the magnetic field with the column density structure. If $\zeta<$ 0.0, the peak of the histogram occurs around $\cos\phi\sim\pm$ 1.0, indicating a perpendicular alignment of the magnetic field with the column density structure. Finally, if $\zeta\sim$0.0, it indicates a flat distribution as a function of relative angle.

\subsubsection{Projected Rayleigh Statistic (PRS)}
As with the HRO method, the detailed methodology of the PRS method is presented elsewhere \citep{Jow2018}. We briefly summarize the key elements of the PRS method here as relevant to our application.

The PRS method determines the angle of relative orientation in a similar manner as the HRO method (Equation \ref{eq:phi}). However, instead of using the histogram shape parameter, the PRS method uses the Rayleigh statistic to determine whether a distribution of angles is uniformly distributed. The Rayleigh statistic, $V'$, is defined as:
\begin{equation}
    V' = \frac{\sum_k^Nw_k\cos(2\phi_k)}{\sqrt{\sum_k^Nw_k^2/2}} \label{eq:V_prime}
\end{equation}
where $\phi$ is the set of relative angles defined using Equation \ref{eq:phi}, $N$ is the number of relative angles in the set, and $w_k$ is the relative weight accounting for the SNR of the polarization measurements, allowing for emphasis of the higher SNR measurements. $V'$ can be viewed as a hypothesis test where $V'\sim$0 indicates that there is no evidence of a preferential angle either parallel or perpendicular between the $N(H_2)$ column density gradient and the magnetic field orientation. A value of the Rayleigh statistic where $V'\gg$0 indicates a significant parallel relative orientation and $V'\ll\,$0 indicates a significant perpendicular relative orientation. 

\subsubsection{Mean Relative Orientation Angle}
An alternative way to evaluate preferential angles is by inspecting the mean relative orientation angle ($\langle\phi\rangle$), defined as:
\begin{equation}
    \langle\phi\rangle = \tan^{-1}\left(\frac{y}{x}\right), \label{eq:phi_mean}
\end{equation}
where $x$ and $y$ are defined such that:
\begin{equation}
    x = \frac{\sum_k^Nw_k\cos(2\phi_k)}{\sum_k^Nw_k},\,\, y = \frac{\sum_k^Nw_k\sin(2\phi_k)}{\sum_k^Nw_k}, \label{eq:xy} 
\end{equation}
where $\phi_k$ and $w_k$ are defined as in Equation \ref{eq:V}. This matches the definition of mean relative angle derived in \citet{Jow2018}.

Generally, $V^\prime$ scales as the square root of the number of measurements. This can be traced to its relationship to the random walk problem \citep{Jow2018}. Because of this, it is important to ensure that only data that are independent count toward the PRS. In calculating $V^\prime$, we have utilized the FIREPLACE pixelization, which subsamples the effective resolution (taken to be that of the column density map.)  The $V^\prime$ statistic must be corrected to account for this oversampling bias.   To do this, we reduce $V^\prime$ as follows

\begin{equation}
    V = \left(\frac{\Delta{}l}{FWHM_{N(H_2)}}\right)\times{}V', \label{eq:V}
\end{equation}
where $\Delta{}l$ is the kernel size used to evaluate the relative orientation (4\farcs9) and $FWHM_{N(H_2)}$ is the resolution of the column density map (25\arcsec). 

\subsubsection{Special Consideration when Comparing to NTF Orientation}


Since the RHT algorithm smooths the MeerKAT data set to an 8\arcsec resolution we use a smoothing kernel of $\sigma\sim$8\arcsec\ (1 pixel) in calculating the gradient. This pixelization corresponds to approximate Nyquist sampling of the FIREPLACE 19\farcs6 beam size, which sets the resolution for the analysis. As in Equation~\ref{eq:V} above, we need to account for the oversampling bias. For this case, the debias factor utilizes the  pixel scale of the filtered MeerKAT data (8\farcs0) in the numerator, and the FIREPLACE angular resolution (19\farcs6) in the denominator.  

The HRO and PRS calculations were performed using the Magnetar code developed by Juan Soler and available on GitHub.

\section{CMZ-wide Magnetic Field Alignment} \label{sec:cmz_align}
\subsection{Field Alignment with Molecular Structure}
\begin{figure*}
    \centering
    \includegraphics[width=1.0\textwidth]{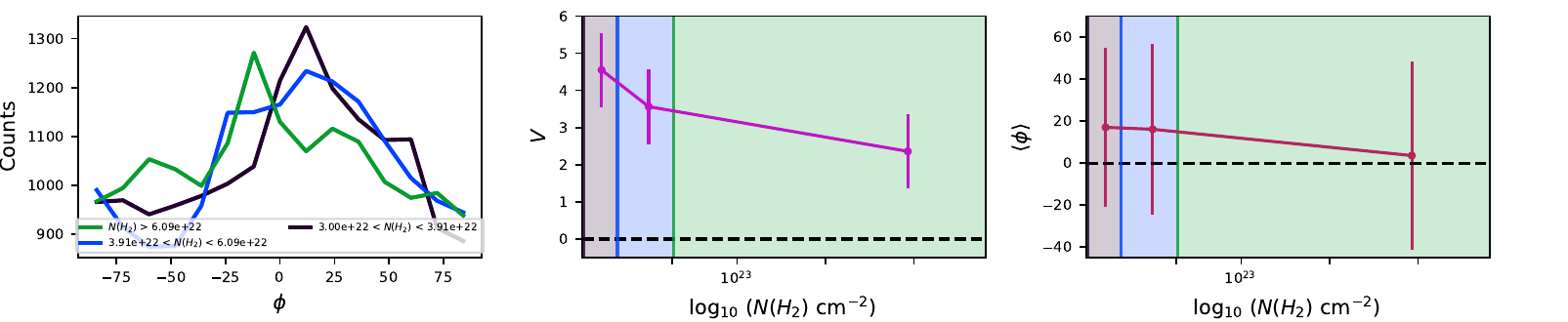}
    \caption{Quantification of the relative alignment between the magnetic field orientation and column density gradient for the entire.  Left: the HROs obtained over three different high column density ranges as marked in the plot. The minimum column density evaluated is marked with the contour in the upper panel of Figure \ref{fig:NH2}.  Middle: the PRS statistic for the different density regimes shown in the HRO panel. Right: the mean angle derived for the density regimes shown in the HRO panel. the background colors in the $V$ and $\langle\phi\rangle$ panels indicate the extents of the column density regimes analyzed.}
    \label{fig:NH2_HRO}
\end{figure*}
We first test whether there is a systematic orientation between the magnetic field derived from FIREPLACE and the molecular structure across the entire CMZ. To do so, we analyze the relative angle in only the higher column density regimes of the CMZ where we choose $N(H_2)>3\times{}10^{22}$ cm$^{-2}$, as marked by the contour level in Figure \ref{fig:NH2}. Column densities of this value and above generally have reduced $\chi^2\leq2.0$. For these higher column densities we compute the relative angle over three distinct column density regimes. These density regimes are dynamically chosen to ensure that an approximately equal number of relative orientations are evaluated within each column density regime. We obtain approximately 16,200 independent relative orientation measurements in each column density bin. This method ensures that the relative orientation found over each density regime is of equivalent statistical significance.

Figure \ref{fig:NH2_HRO} displays the relative alignment of the magnetic field with the gradient of the column density distribution derived from Herschel. The panels, from left to right, display the HROs calculated for the different column density regimes, the PRS $V$ statistic, as a function of column density, and the mean relative angle observed in each column density regime. We do not present the HRO shape parameter, $\zeta$, since $V$ conveys this information more completely. 

From this plot we observe that the field has a strong preference for parallel orientation to the column density structure (middle panel of Figure \ref{fig:NH2_HRO}). The highest density region analyzed corresponds to the brightest and most dense molecular clouds within the CMZ like Sgr B2, the 20 \kms\ cloud, and the Brick. The locations of these clouds within the CMZ are marked in Figure \ref{fig:legend}. The lower column density regimes trace lower density clouds within the CMZ. 

The observed preference for parallel orientation at high column density agrees with the preliminary magnetic field orientation analysis conducted in \citetalias{Pare2024} that indicated the magnetic field in the brightest $I_{214}$ intensity regimes of the CMZ are more aligned with the Galactic Plane. This trend in magnetic field was demonstrated in the histograms shown in their Figures 13 and 14. The result found in this work is consistent with the trend observed in \citetalias{Pare2024} indicating the molecular structures are likely sheared in the direction parallel to the Galactic plane.

\subsection{Field Alignment With the NTFs}
\begin{figure}
    \centering
    \includegraphics[width=0.45\textwidth]{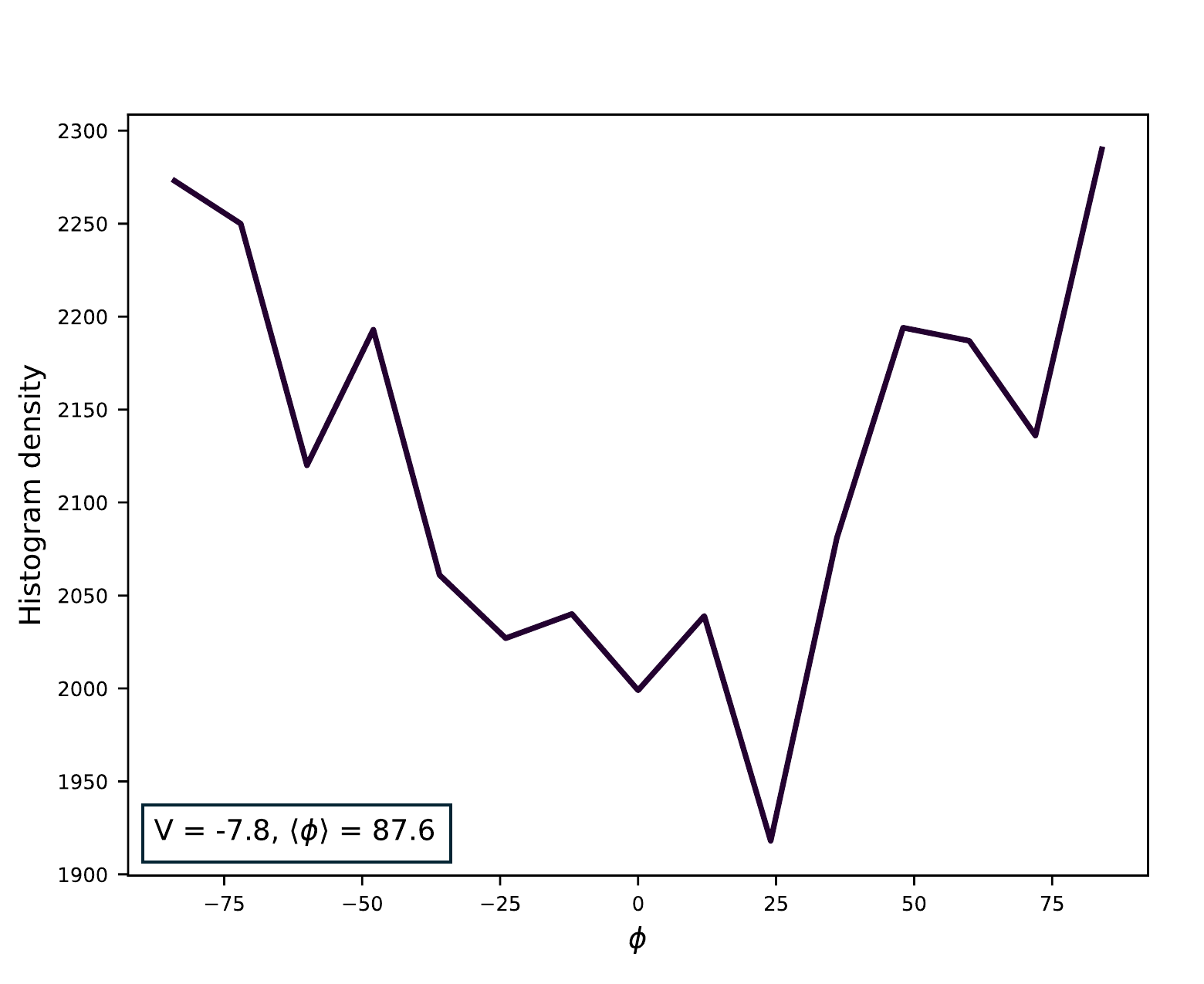}
    \caption{Quantification of the relative alignment between the magnetic field orientation and the NTF orientation. The figure shows the HRO obtained over the RHT significance range sensitive to the emission originating from the NTFs shown in Figure \ref{fig:MeerKAT_NTF}. The PRS $V$ and $\langle\phi\rangle$ values are shown in the lower-left corner of the figure.}
    \label{fig:NTF_HRO}
\end{figure}
The HRO comparison of the magnetic field and NTF orientation is shown in Figure \ref{fig:NTF_HRO}. We obtain roughly 7900 independent relative orientation measurements to conduct this statistical analysis. The relative alignment between the FIREPLACE magnetic field and the NTFs reveals a generally perpendicular relative orientation, $V\ll0$. The mean relative angle is found to be $\langle\phi\rangle\sim$88\degree. There are filamentary features inferred from the RHT that are not associated with these NTFs. Since these non-NTF filaments have lower RHT significance, we only evaluate the relative alignment of the magnetic field for RHT significance $>$0.15. The perpendicular relative orientation indicates that the FIREPLACE magnetic field measurements that coincide with the NTFs are largely oriented perpendicular to the NTF orientation. This anti-correlation has been observed for discrete sources local to the Radio Arc. For example Kuiper Airborne Observatory (KAO) 60 \micron\ observations of the Sickle \hii\ region reveal magnetic field measurements that are perpendicular to the orientation of the Radio Arc filaments \citep{Dotson2000,Chuss2003a}.

Preliminary magnetic field direction results in \citetalias{Pare2024} indicated a bimodal magnetic field distribution where one component is aligned perpendicular to the Galactic plane which is seemingly aligned with the NTF orientation. The fact that the HRO/PRS results show a perpendicular orientation to the NTFs indicates that the FIREPLACE-derived magnetic field measurements that are aligned with the NTFs do not universally coincide with these non-thermal structures.

\section{Magnetic Field Alignment Within Individual CMZ Clouds and NTFs} \label{sec:cloud_align}
We now determine whether the CMZ-wide relative angle trend seen in Figure \ref{fig:NH2_HRO} is also observed within individual molecular clouds in the region. To do so, we select prominent CMZ clouds located throughout the region that have at least 100 FIREPLACE magnetic field measurements in each column density bin that coincide with the cloud. The locations and names of the selected molecular clouds within the CMZ are shown in Figure \ref{fig:legend}. We also inspect the relative orientation between the FIREPLACE magnetic field and prominent NTFs in the CMZ that have a significant number of coincident polarization measurements. We summarize the alignment results in Table \ref{tab:cloud_param}, where we show the average $V$ and $\langle\phi\rangle$ for each cloud. We also display the peak $N(H_2)$ column density and star formation rates (SFRs) for the prominent molecular clouds.

\subsection{Alignment in Prominent Molecular Clouds}
\subsubsection{The Ring Cloud}
\begin{figure*}
    \centering
    \includegraphics[width=1.0\textwidth]{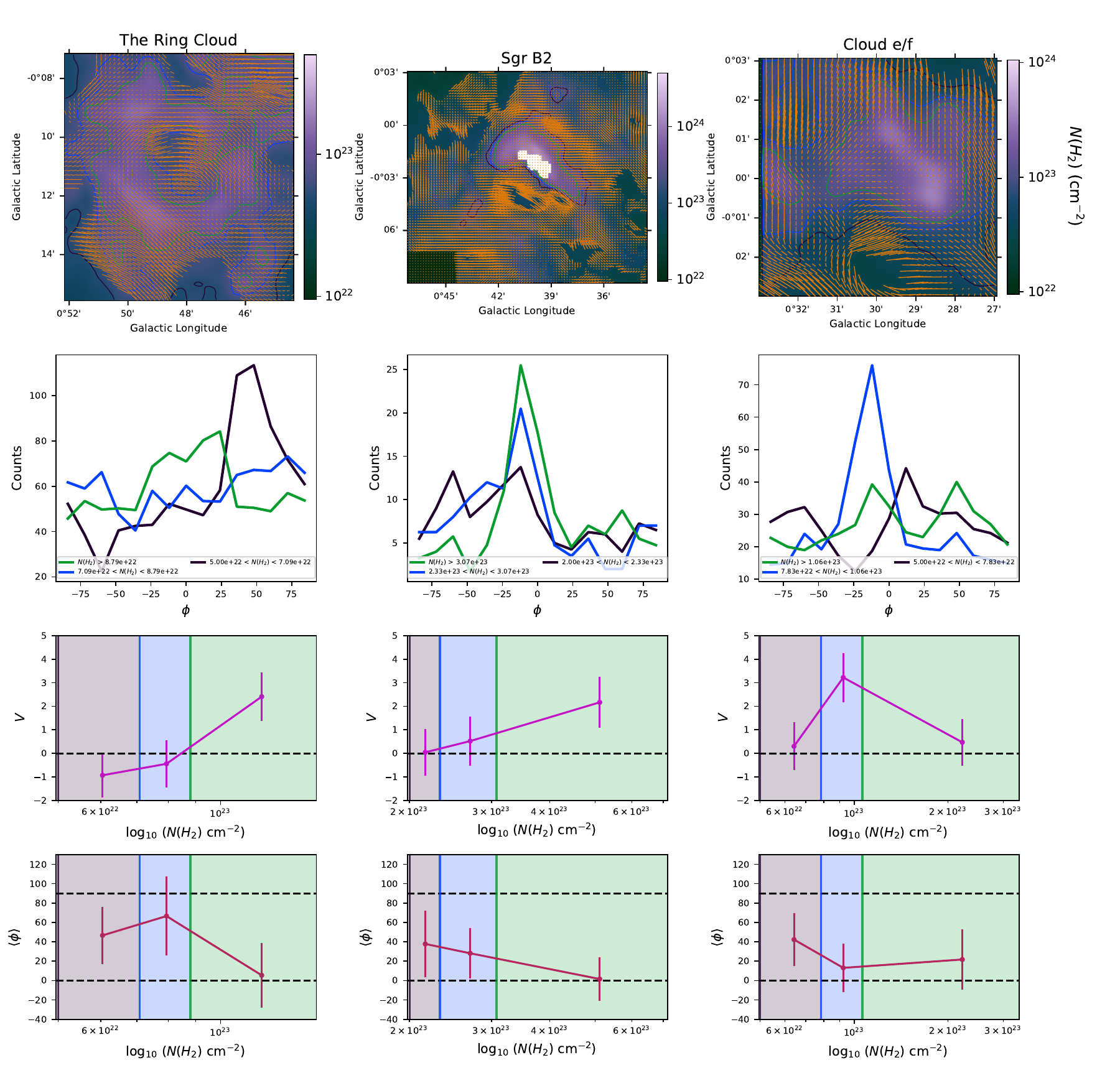}
    \caption{HRO and PRS results obtained for (from left to right) the Ring cloud, Sgr B2, and Cloud e/f. Top row: the column density distribution for each cloud derived from Herschel is shown with FIREPLACE pseudovectors overlaid. The contours indicate the column density regimes over which the HRO and PRS statistics are determined. Second row: the HROs determined for the three column density regimes indicated in the plots. Third row: the PRS Rayleigh statistic calculated for each column density regime. Bottom row: the mean relative angle calculated for each column density regime is shown. The background colors in the $V$ and $\langle\phi\rangle$ panels indicate the extents of the column density regimes analyzed.}
    \label{fig:multi1}
\end{figure*}
We first analyze the Ring cloud (M0.8-0.2) observed to the Galactic Southeast of Sgr B2 (Figure \ref{fig:legend}). This cloud has been identified previously and is thought to be a wind-blown shell \citep[][hereafter referred to as FIREPLACE II]{Butterfield2024b}. This structure was called the M0.8-0.2 Ring in that paper and in \citet{Nonhebel2024}; however, in this work we will simplify this nomenclature to ``the Ring cloud.''

For our analysis of the Ring cloud we take our minimum column density as 5$\times$10$^{22}$ cm$^{-2}$ and assess the relative alignment between the magnetic field and the column density structure over three column density regimes. As for the approach for the entire CMZ, these column density regimes are dynamically determined such that each bin has approximately the same number of relative orientations. This method is also employed for all the clouds discussed later in this section. The number of relative orientations per bin is different for each cloud, with approximate numbers for the independent measurements per bin shown in the fourth column of Table \ref{tab:cloud_param}. We note that there are instances where the column density regimes are less than a beam size apart, so there is possible correlation between the different regimes.

The HRO and PRS relative alignment results for the Ring cloud are shown in the left column of Figure \ref{fig:multi1}. In this column, the topmost panel shows the Ring cloud column density distribution derived from Herschel with the FIREPLACE derived magnetic field directions overlaid as orange pseudovectors. The contours indicate the column density levels over which the HRO and PRS algorithms were run, with the colors of the contours matching the colors used for the HROs. The remaining panels, in descending order, show the HRO results for the three different column density regimes, colored in agreement with the contour level colors; the PRS $V$ statistic (Equation \ref{eq:V}), as a function of column density; and the mean relative angle $\langle\phi\rangle$ (Equation \ref{eq:phi_mean}) as a function of column density. As with the CMZ-wide HRO results we do not present the HRO shape parameter, $\zeta$, since it is redundant with $V$.

The Ring cloud exhibits perpendicular orientation in the lowest  column density regime, no preferential alignment in the middle column density regime, and then becomes parallel at the highest column density regime. This is shown by the $V$ distribution for the Ring cloud, which is displayed in the third panel in the left-hand column of Figure \ref{fig:multi1}. $V\sim-1$ at the lowest column density regime, $V\sim0$ in the middle column density regime, and $V\sim2$ at the highest column density regime. Our $\langle\phi\rangle$ distribution for this cloud is $\sim$40\degree\ for the lowest column density regime, increases to $\sim$60\degree\ in the middle column density regime, and drops to $\sim$5\degree\ for the highest column density regime.

\subsubsection{Sgr B2}
Sgr B2 is located at the Easternmost end of the CMZ and is the most massive molecular cloud in the entire Galaxy. It exhibits significant star formation \citep{Schmiedeke2016,Ginsburg2018,Pan2024}. To analyze the relative angle between the magnetic field and cloud morphology within this cloud complex, we limit our analysis to the highest column density regions of Sgr B2, corresponding to the Sgr B2 core \citepalias{Butterfield2024a}. This core region of Sgr B2 is where the majority of the star formation within the cloud complex is occurring \citep{Belloche2013}. 

For our analysis of Sgr B2 we take our minimum column density as 2$\times$10$^{23}$ cm$^{-2}$ and assess the relative alignment between the magnetic field and the column density structure over three column density regimes. We note that some lines of sight in Sgr B2 are masked out because the high column densities in that region are not well fit by our model. The HRO and PRS relative alignment results for Sgr B2 are shown in the middle column of Figure \ref{fig:multi1}, where the ordering of the panels in this column matches that for the Ring cloud.  

The results obtained for the Sgr B2 core region reveal that Sgr B2 has V$\sim$0 for the lowest density region.  The PRS Rayleigh statistic, $V$, indicates an increasingly preferential parallel orientation as column density increases, with $\langle\phi\rangle$ ranging from 40 -- 0\degree. 

\subsubsection{Cloud e/f}
Cloud e/f is part of the larger dust ridge complex that is comprised of multiple molecular clouds located between Sgr B2 and the Brick \citep{Longmore2013}. The clouds within this dust ridge are suggested to be Young Massive Cluster (YMC) precursors that are dense enough to be forming stars but are generally observed to be quiescent \citep{Lis1999,Rathborne2015,Walker2015,Walker2018}. Cloud e/f is on the Eastern edge of the dust ridge and is the brightest of these precursors at infrared wavelengths other than the Brick, which we discuss later.

We present the HRO and PRS statistics derived for Cloud e/f in the right column of Figure \ref{fig:multi1}, where the format of the panels in this column matches that of the Sgr B2 column. The minimum column density evaluated in this analysis is $5\times10^{22}$ cm$^{-2}$. All clouds that follow in this section use this same threshold column density to determine the HRO/PRS results. For Cloud e/f, analysis of the relative angle between the magnetic field and the cloud structure reveals a preference for a parallel magnetic field orientation. This trend can be clearly observed in the PRS Rayleigh statistic, $V$, panel (third from the top). $V\sim0$ for the lowest column density regime, approaches $V\sim3$ for the middle column density regime (indicating a strongly parallel relative angle), and then $V\sim1$ for the highest column density regime.

Further supporting this trend is the plot of mean relative angle, $\langle\phi\rangle$, shown in the bottom panel of the right-hand column of Figure \ref{fig:multi1}. The mean angle is close to 40\degree\ at the lowest column density and approaches 20\degree\ for the intermediary and highest column density regimes.

\subsubsection{The Bricklets}
\begin{figure*}
    \centering
    \includegraphics[width=1.0\textwidth]{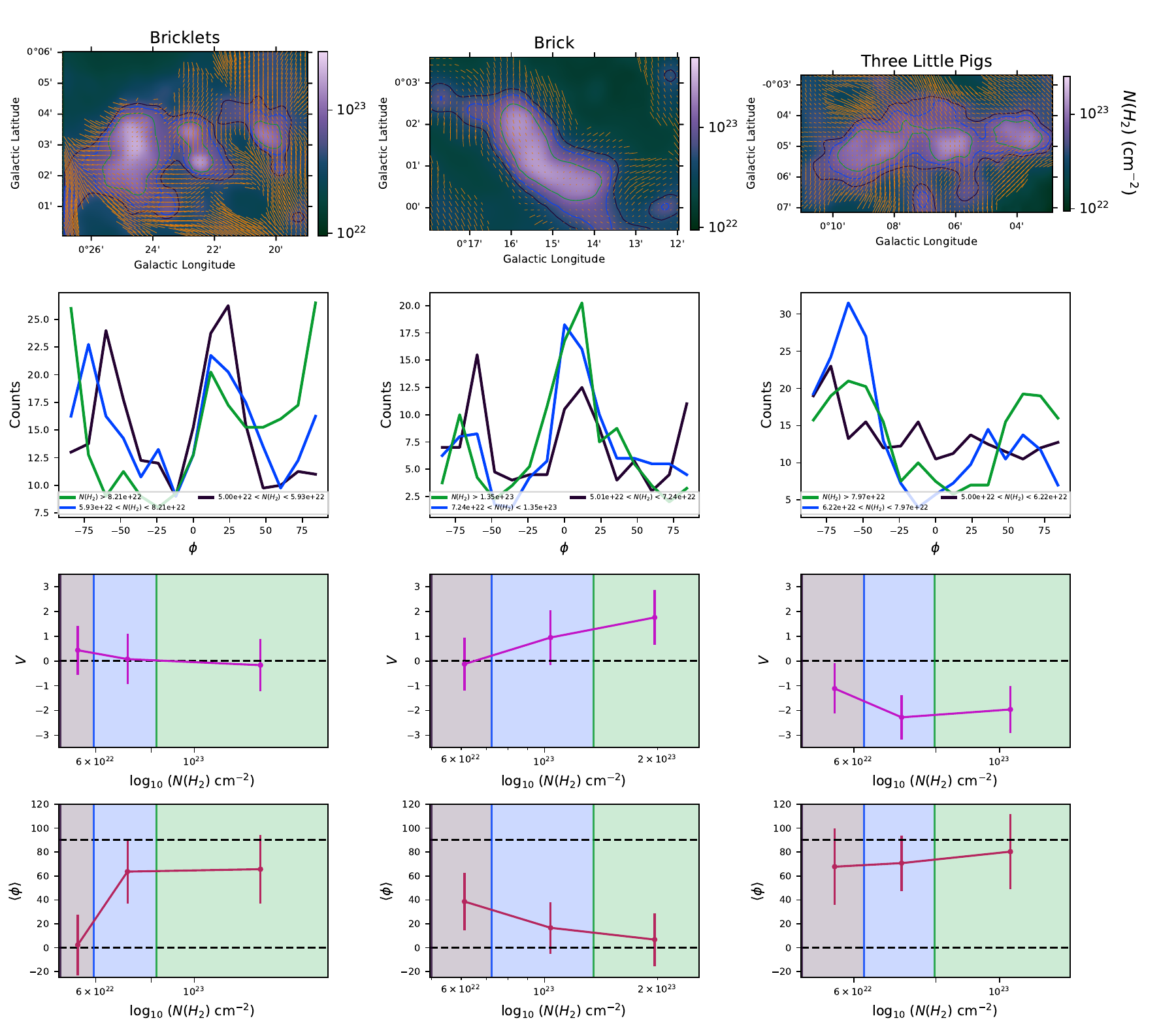}
    \caption{The HRO and PRS results obtained for the Bricklets, the Brick, and the TLP cloud complex. The panel format is the same as for Figure \ref{fig:multi1}.}
    \label{fig:multi2}
\end{figure*}
We use the term ``Bricklets'' to refer to clouds b, c, and d within the Dust Ridge \citep{Krieger2017}. These clouds are located between the Brick and Cloud e/f \citep{Longmore2013,Walker2015,Rathborne2016,Walker2018,Battersby2020}. Though these clouds are likely YMCs, only cloud c is observed to be actively forming stars \citep{Ginsburg2015,Walker2018}.

We present the HRO and PRS results for the Bricklets in the left column of Figure \ref{fig:multi2}. This set of clouds has small $V$ values, that indicate an increasingly perpendicular relative orientation as column density increases. Here, $V\sim0.5$ at the lowest of the three column density regimes, with  $V\sim0$ in the middle column density regime, and slightly negative at the highest column density regime. We observe a notable transition in $\langle\phi\rangle$ from a 0\degree\ relative alignment in the lowest column density regime to a 60\degree\ relative alignment for the highest column density regime.

\subsubsection{The Brick}
The Brick is a high density cloud that is located at the Western end of the dust ridge \citep{Lis1999} and, as with the Bricklets, it exhibits little  evidence of active star formation \citep{Longmore2012,Mills2015,Lu2019,Walker2021}. This cloud is close in projection to Sgr A$^*$ and is proximal to the Radio Arc NTF \citep{Yusef-Zadeh1987,Lang1999a}. The analysis of the orientation between the magnetic field and cloud structure obtained from both the HRO and PRS methods is shown in the middle column of Figure \ref{fig:multi2}.

The Brick displays an increasingly parallel magnetic field orientation with increasing column density. This increasingly parallel orientation can be observed in the HRO results for this cloud where the higher density histograms (represented by the lighter-colored histograms) have peaks close to 0\degree. Furthermore, $V$ is increasingly positive as a function of column density, indicative of an increasingly significant parallel relative angle. Finally, we see that $\langle\phi\rangle$, shown in the bottom panel, approaches 0\degree\ as the column density increases.

\subsubsection{The Three Little Pigs}
The Three Little Pigs (TLP) cloud complex is comprised of three molecular clouds known as M0.145-0.086 (the Straw), M0.106-0.082 (the Sticks), and M0.068-0.075 \citep[the Stone,][]{Battersby2020}. The close proximity of these clouds, and the fact that they all have a similar velocity around 50 \kms\ indicates that these clouds are likely dynamically connected and not just close in projection \citep{Kruijssen2015,Butterfield2018}. We therefore treat all of the clouds in this complex together in our analysis. The results obtained from both the HRO and PRS methods is shown in the right column of Figure \ref{fig:multi2}.

Unlike the results obtained for the full CMZ, the Sgr B2 core, and the Brick, the relative angle between the column density structure and magnetic field orientation for the TLP becomes increasingly orthogonal to the cloud as a function of column density. This trend is shown in the plot of $V$ for which the value becomes increasingly negative with increasing column density, indicating an increasingly significant perpendicular relative orientation. We see that the mean relative angle, $\langle\phi\rangle$, in the bottom panel increases from 60\degree\ to 80\degree\ as the column density increases.

\subsubsection{The 20 km/s Cloud}
\begin{figure*}
    \centering
    \includegraphics[width=1.0\textwidth]{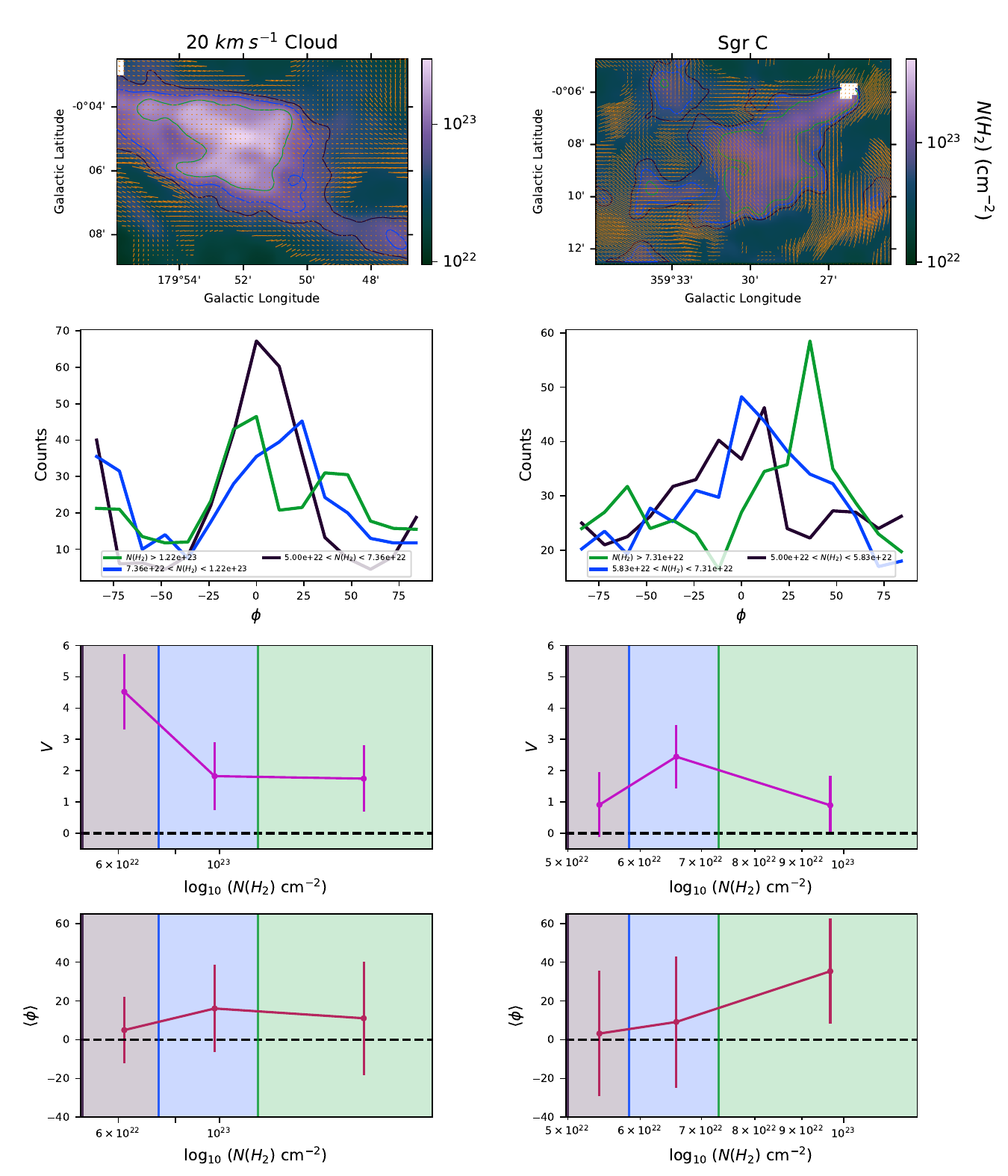}
    \caption{The HRO and PRS results obtained for the 20 \kms\ cloud and Sgr C. The panel format is the same as for Figure \ref{fig:multi1}.}
    \label{fig:multi3}
\end{figure*}
The 20 \kms\ cloud is close in projection to the central black hole of the Galaxy, Sgr A$^*$, and is thought to be in an early stage of star formation \citep{Bally2010,Lu2015}. The highest column density in this central 50 pc region coincides with the 20 \kms\ cloud.

The HRO and PRS results for this cloud is shown in the left column of Figure \ref{fig:multi3}. This cloud exhibits a strongly parallel relative orientation for all column density regimes. The mean relative orientation is close to 0\degree\ for all these column density regimes studied as shown in the bottom panel of the column. Here $V>0$, further indicating a strong preference for a systematically parallel orientation.

\subsubsection{Sgr C}
The Sgr C molecular cloud is located in the far West of the CMZ, and it possesses the second highest rate of star formation in the region besides that of Sgr B2 \citep{Kendrew2013}. Sgr C is comprised of a high density head-tail structure (located at G359.44-0.102) and diffuse dust that surrounds this more compact, high density structure \citep{Lu2019,Liu2024}. The HRO/PRS statistic results are shown in the right column of Figure \ref{fig:multi3}.

As with the 20 \kms\ cloud, the $V\geq1$ for Sgr C is universally positive for all column density regimes. The mean relative angle, $\langle\phi\rangle$, is close to 0\degree\ at lower column densities but does approach 40\degree\ for the highest column density regime.

\begin{table*}
\begin{threeparttable}
\caption{CMZ Cloud \& NTF Properties}
\begin{tabular}{cccccc}
\hline
Cloud Name & Average $V$ & Average $\langle\phi\rangle$ & \# Measurements & Peak $N(H_2)$ (cm$^{-2}$) & SFR$^*$ (10$^{-3}$ M$_{\odot}$ yr$^{-1}$) \\ \hline\hline
The Ring Cloud & 0.3 & 40\degree\ & 880 & 1.5$\times$10$^{23}$ & NA \\
Sgr B2 & 0.9 & 23\degree\ & 160 & 1.9$\times$10$^{24}$ & 62.0 \\
Cloud e/f & 1.3 & 26\degree\ & 400 & 3.4$\times$10$^{23}$ & 16.9$\pm$10.1 \\
The Bricklets & 0.1 & 44\degree\ & 230 & 2.0$\times$10$^{23}$ & 38.7$\pm$23.2 \\
Brick & 0.8 & 21\degree\ & 110 & 2.6$\times$10$^{23}$ & 2.3$\pm$1.4 \\
TLP & -1.8 & 73\degree\ & 210 & 1.3$\times$10$^{23}$ & 3.0$\pm$1.0 \\
20 \kms\ Cloud & 2.7 & 11\degree\ & 350 & 2.9$\times$10$^{23}$ & 6.3$\pm$3.8 \\
Sgr C & 1.4 & 16\degree\ & 440 & 1.7$\times$10$^{23}$ & 65.4$\pm$39.2 \\
\hline
Radio Arc & -1.1 & 89\degree\ & 960 & NA & NA \\
Sgr C NTF & -1.2 & 83\degree\ & 170 & NA & NA \\
\hline
\end{tabular} 
\begin{tablenotes}
\small
\item * SFR values and errors are from \citet{Ginsburg2018} for Sgr B2 and \citet{Hatchfield2023} for all other clouds. The Ring cloud does not currently have any star formation estimates.
\end{tablenotes}
\label{tab:cloud_param}
\end{threeparttable}
\end{table*}

\subsection{Alignment in Prominent NTFs}
\subsubsection{The Radio Arc}
\begin{figure}
    \centering
    \includegraphics[width=0.32\textwidth]{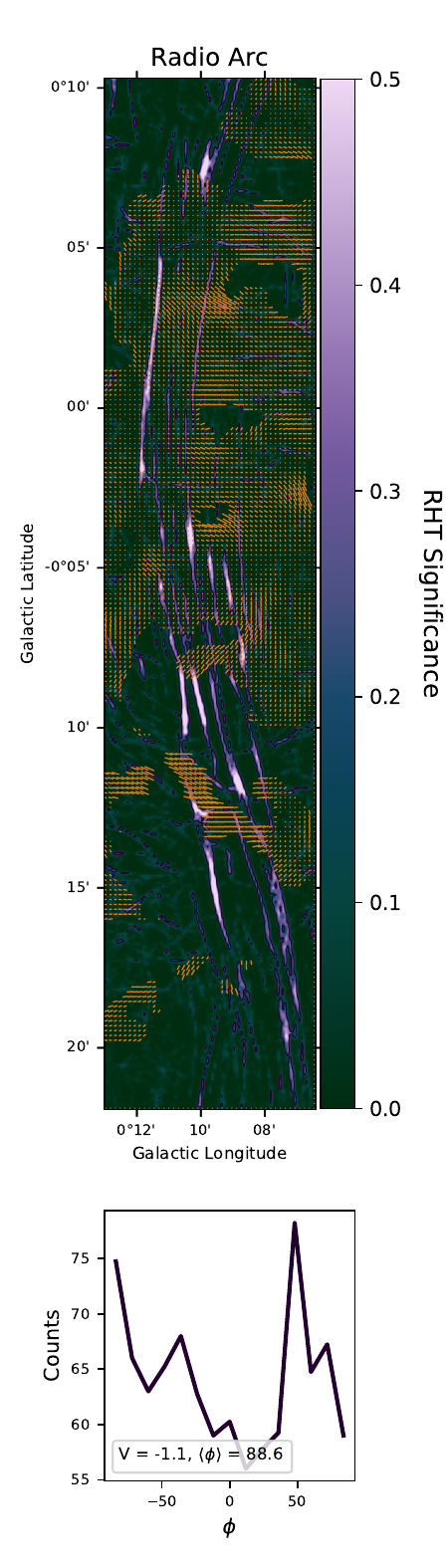}
    \caption{The HRO and PRS results obtained for the Radio Arc NTF. The top panel shows the filtered MeerKAT 1.28 GHz intensity distribution. The bottom panel shows the HRO determined from this filtered distribution. The PRS Rayleigh statistic ($V$) and $\langle\phi\rangle$ values determined from this region are shown in the lower left corner of this panel. The magnetic field inferred from dust polarimetry tends to be perpendicular to the filament orientation.}
    \label{fig:NTF_multi1}
\end{figure}
The Radio Arc is the brightest and largest (in the plane of the sky) of the NTFs in the GC and has been studied for 40 years \citep[e.g.,][]{YMC1984,Yusef-Zadeh1987,Pare2021,Yusef-Zadeh2022,Pare2024b}. The Radio Arc may be interacting with the Sickle \hii\ region, which is thought to be the ionized edge of a molecular cloud \citep{Serabyn1991}. There is also evidence of interaction between the Radio Arc and the arched filament molecular structures based on the morphological changes observed at the interaction points for both the NTF and arched filaments \citep{YMC1984,Morris1989}.
 
For the purpose of the analysis presented in this paper we therefore assume the dust and Radio Arc are interacting. Since there are also a large number of FIREPLACE polarization measurements ($\sim$1000) that coincide with the Radio Arc we study how the relative angle of the magnetic field coincident with the Radio Arc compares to what is observed for the larger NTF population in Figure \ref{fig:NTF_HRO}. Our relative angle analysis for the Radio Arc is presented in Figure \ref{fig:NTF_multi1}.

We observe a generally perpendicular relative orientation for the radio intensity regime studied, with $V<$0. We derive a $\langle\phi\rangle\sim$89\degree. The generally perpendicular relative angle observed for the Radio Arc agrees with what is observed for the larger NTF population in Figure \ref{fig:NTF_HRO}.

\subsubsection{The Sgr C NTF}
\begin{figure}
    \centering
    \includegraphics[width=0.43\textwidth]{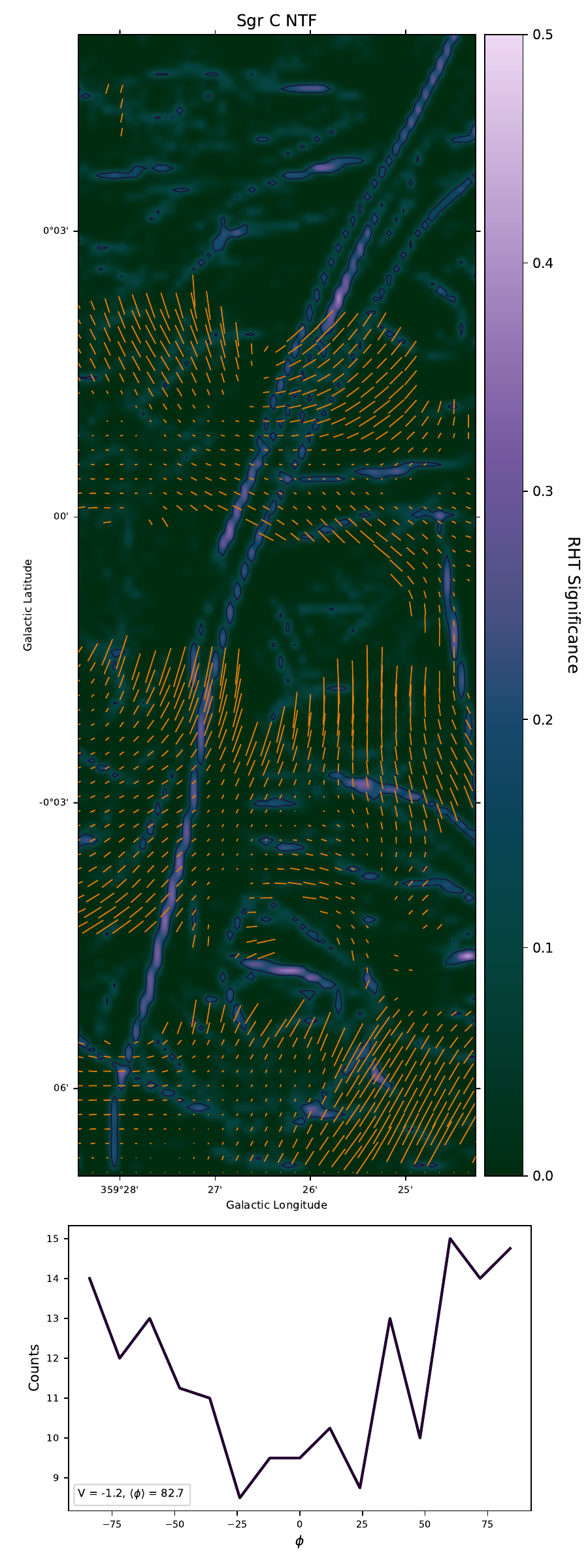}
    \caption{The HRO and PRS results obtained for the Sgr C NTF. The format of the panels in this figure is the same as for Figure \ref{fig:NTF_multi1}.}
    \label{fig:NTF_multi2}
\end{figure}
We also study an NTF near the Sgr C molecular cloud. This is also a larger NTF, and like the Radio Arc it coincides with a significant number of FIREPLACE magnetic field measurements ($\sim$100). Furthermore, there is morphological evidence of potential interaction between this NTF and the coincident molecular structures. The NTF bifurcates and bends at locations along its length that are coincident with molecular structures, and it is thought to be co-spatial with Sgr C molecular structures based on HI absorption observations \citep[e.g.,][]{Lang2010,Roy2003}. These studies reported changes in HI absorption observed along the length of the Sgr C NTF, indicating that the NTF is at least partially embedded within a local molecular cloud. Besides the 1.28 GHz MeerKAT observations \citep{Heywood2022}, this NTF has been studied at 6\farcs5 resolution using the Giant Meterwave Radio Telescope \citep{Roy2003} and at 15\arcsec\ resolution using the Very Large Array \citep{Lang2010}. As with the Radio Arc, we compare how the relative angle between the magnetic field and the Sgr C NTF compares with what is observed for the larger NTF population in Figure \ref{fig:NTF_HRO}. Our relative angle analysis for the Sgr C NTF is presented in Figure \ref{fig:NTF_multi2}.

As with the Radio Arc, we observe a generally perpendicular relative orientation between the filament and the magnetic field inferred from dust polarization for the Sgr C NTF in Figure \ref{fig:NTF_multi2}. The Rayleigh statistic value is $V<$0 and a mean angle of $\langle\phi\rangle\sim$83\degree. As with the Radio Arc, this generally perpendicular orientation agrees with what is observed for the larger NTF population as shown in Figure \ref{fig:NTF_HRO}.

\section{DISCUSSION} \label{sec:disc}
\subsection{Analysis of Relative Orientation Within CMZ Clouds} \label{sec:cmz_cloud_align}
\begin{figure}
    \centering
    \includegraphics[width=0.45\textwidth]{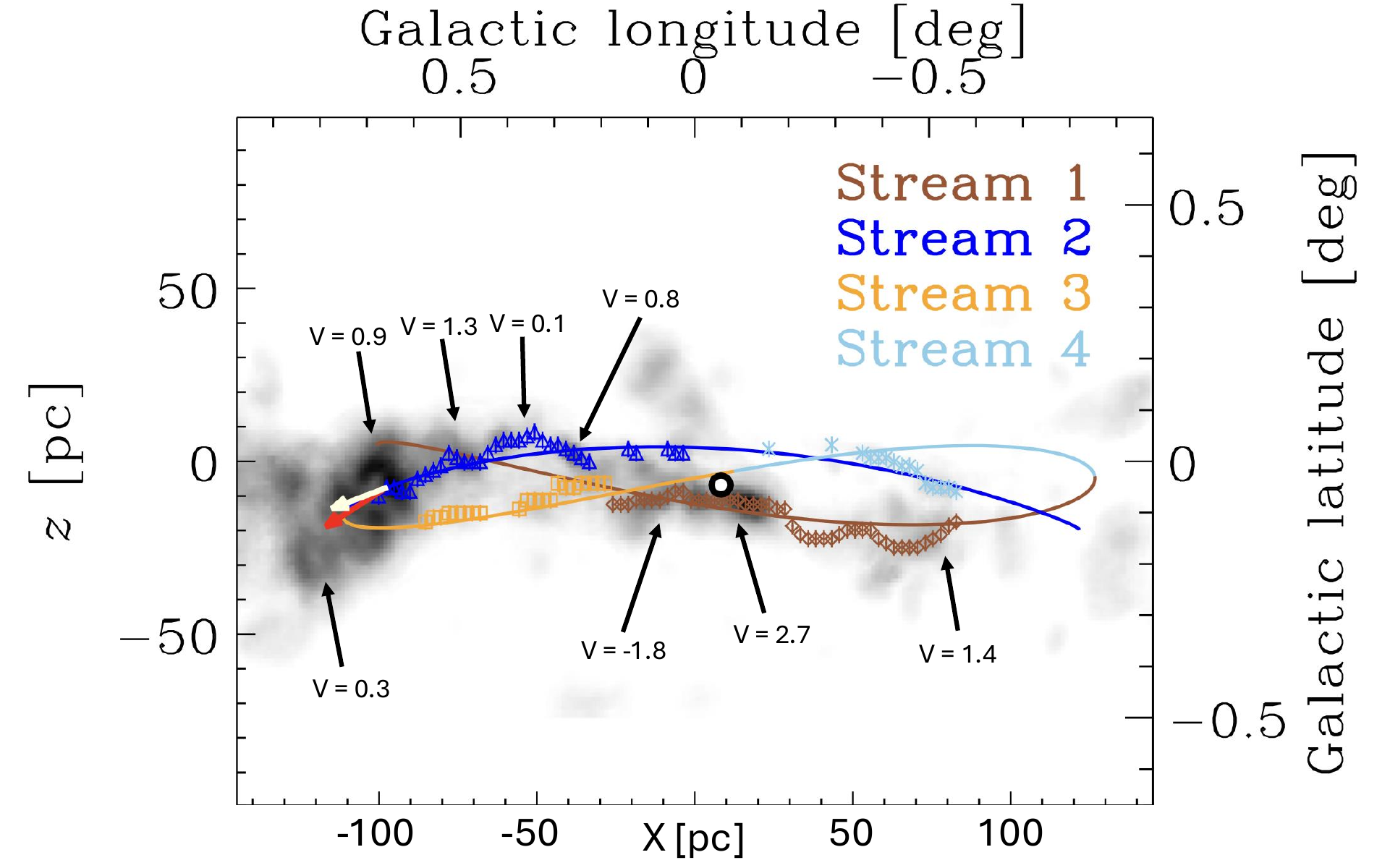}
    \caption{The 4-stream model from \citet{Kruijssen2015} (top panel of Figure 4 of that work) with the average $V$ values found in this work derived for prominent CMZ clouds overlaid.}
    \label{fig:Kruijssen2015}
\end{figure}
There is a significant variation in the CMZ cloud relative orientation observed in Figures \ref{fig:multi1} -- \ref{fig:multi3}. Some of the clouds exhibit similar relative alignment trends to what is observed for the full CMZ. Specifically, Sgr B2, the Brick, the 20 \kms\ cloud, and Sgr C all demonstrate ubiquitously positive $V$ values in all column density regimes.

The remaining clouds, however, show markedly different relative alignment trends. While the Ring cloud exhibits $V<0$ for intermediate column densities ($N(H_2)\sim10^{23}$ cm$^{-2}$), Cloud e/f has a strong $V>0$ signature at intermediate column densities. In comparing the Brick and TLP clouds, the clouds show opposite trends as the column density increases, with the Brick having $V>0$ and the TLP having $V<0$. This difference could be due to variations in either the environments or the stage of dynamical evolution of the clouds. 

We also compare the relative orientation results with the 4-stream model for the CMZ cloud geometry developed in \citet{Kruijssen2015}. In their model, the Brick is associated with stream 2 which has a LOS velocity of 20 \kms, whereas the TLP is part of a distinct stream (stream 1) moving with an LOS velocity of 50 \kms\ \citep[Figure 4 of][]{Kruijssen2015,Butterfield2018}. This figure from \citet{Kruijssen2015} is reproduced in this work and is shown in Figure \ref{fig:Kruijssen2015}, where we have overlaid the average $V$ values derived for the prominent CMZ cloud studied in this work.

The range of alignment trends observed for CMZ clouds leads to the question of whether there is any dependence on magnetic field alignment with molecular cloud properties or location within the CMZ. Table \ref{tab:cloud_param} lists the alignment results for the prominent CMZ clouds presented in this work. Also shown are the peak column densities for each cloud derived from the column density distribution shown in Figure \ref{fig:NH2} and the SFRs derived for the clouds in previous works \citep{Ginsburg2018,Hatchfield2023}. We see a general trend of higher $V$ for the clouds that have the largest peak column density values (i.e. Sgr B2, cloud e/f, the 20 \kms\ cloud, and the Brick). Furthermore, the cloud that has significant $V<0$ is the TLP which has the lowest lowest peak column density. We see no obvious correlation between the cloud SFRs and the magnetic field alignment.

We also compare the magnetic field alignment with the locations of the clouds in the CMZ, which are shown in Figure \ref{fig:legend}. The distribution of $V$ values associated with these clouds is also shown in Figure \ref{fig:Kruijssen2015}. There is no obvious pattern of $V$ with respect to the positions of the clouds in the CMZ. If the clouds are moving along molecular streams, as posited by \citet{Kruijssen2015}, it is possible to also estimate the orbital velocities of the CMZ clouds. This estimate is performed in \citet{Kruijssen2015}, and they derive generally higher orbital velocities in the central region of the CMZ ($\sim$180 -- 200 \kms, corresponding to Cloud e/f, the Bricklets, The Brick, and the 20 \kms\ cloud) than at the edges of the CMZ (120 -- 150 \kms, corresponding to Sgr B2 and Sgr C). There is therefore no clear correlation between the estimated orbital velocity from \citet{Kruijssen2015} and the magnetic field alignment observed in this work.

One other aspect of the molecular clouds to consider is the presence of shocks. Shocks are undoubtedly important in the CMZ molecular cloud population, and several of the prominent cloud structures studied in this work are subject to shocks. For example, \citet{Pillai2015} argue for a shock influencing the magnetic field observed for the Brick. A possible source of this shock is an expanding bubble identified in \citet{Henshaw2022} that is coincident with the Brick. For the other molecular structures observed in this work, the TLP cloud complex and Sgr C exhibit evidence of shocks through interaction with shell-like structures, the orbital motion of the clouds, and interaction with prominent \hii\ regions \citep{Lang2010,Hankins2020,Butterfield2022,Pare2024}. The Ring cloud is also likely a shocked structure, since \citetalias{Butterfield2024b} identify the Ring cloud as a structure generated from expanding winds. We observe a range of magnetic field orientations corresponding to these particular molecular structures.

\subsection{Comparison to Relative Angle Trends in the Galactic Disk}
The orientation between the magnetic field and molecular cloud structure has previously been studied in Galactic Disk clouds. A population of star-forming regions was analyzed in \citet{PlanckXXXV} where the clouds were located at a distance of $\sim$450 pc from the Earth. These clouds are 18$\times$ closer to Earth than the clouds in the GC ($\sim$8.2 kpc away). We can compare this Galactic Disk cloud population with the CMZ clouds since the physical scales sampled by a beam are similar in each case. Specifically, the angular resolution of the CMZ observations is 19\farcs6 (corresponding to a physical resolution of $\sim$0.78 pc) and the angular resolution of the Galactic Disk observations from Planck studied in \citet{PlanckXXXV} is 5\arcmin\ (corresponding to a physical resolution of $\sim$0.65 pc for these disk clouds). 

The Planck observations do not significantly spatially filter the large-scale dust emission from the diffuse ISM, whereas the FIREPLACE observations do. Since the Galactic disk clouds are relatively close to Earth, the diffuse dust emission is negligible and the Planck observations are dominated by the cloud emission. For the FIREPLACE observations the cloud emission is within the spatial frequencies that the FIREPLACE observations are sensitive to. The filtering of large-scale emission therefore isolates the CMZ cloud emission. Nonetheless, we cannot rule out the possibility that the differences in spatial filtering between the SOFIA/HAWC+ and Planck observations could impact the comparison presented here.

The range of relative angles obtained for CMZ clouds contrasts with the trend observed in Galactic Disk molecular clouds where the relative angle is observed to consistently become less parallel with increasing column density \citep[e.g.,][]{PlanckXXXV}. \citet{PlanckXXXV} and \citet{Soler2019} interpret the perpendicular relative orientation in their clouds to indicate regions where gravity dominates the magnetic field (where they identify a column density threshold of $\log_{10}\,N(H_2) \lesssim 21.7$ cm$^{-2}$ below which the magnetic field dominates). We observe the opposite orientation trend: a generally parallel orientation, even at high column densities. There are two possibilities for the parallel orientation. 1) The GC region could be magnetically dominated (similar to the low density regions studied in \citet{PlanckXXXV}) such that the cloud motion is guided by the magnetic field lines. 2) The magnetic field may be sheared into an orientation along the cloud direction of orbital motion (and hence parallel to the cloud elongation). Because of both the high densities involved and the fact that the magnetic field, cloud elongation, and direction of cloud motion are nearly the same for the cloud streams in the region \citep{Kruijssen2015}, we highly favor the latter. We posit that an initial field is sheared into a field oriented parallel to the orbital motions of the cloud \citep[e.g.,][]{Uchida1985,Chuss2003a}. In light of this picture, the TLP cloud complex exhibits convincing evidence of early-stage shear resulting from the orbital motion of the cloud complex \citepalias{Pare2024}.

\subsection{Comparison of Magnetic Field Alignment with NTF Orientation}
Contrasting with the range of relative orientations we observe for the CMZ molecular clouds, the relative orientations for the prominent NTFs are consistently perpendicular to the magnetic field directions inferred from dust polarization. We note that though the Radio Arc and Sgr C NTF are likely local to the CMZ molecular structures, this may not be true for all of the NTFs in the GC. The perpendicular orientation could therefore be an indication that the NTFs are largely not associated with the molecular structures in the CMZ from which the FIREPLACE magnetic field is derived.

However, there is evidence that the FIREPLACE magnetic field recovers a similar vertical field to that traced by the GC NTFs. \citetalias{Pare2024} observed two magnetic field orientation enhancements occurring at  $\sim\,$-20\degree\ and $\sim\,$-80\degree\ angles relative to the Galactic Disk. The -20\degree\ component could connect to the molecular clouds within the CMZ. The -80\degree\ component, however, could be the vertical field inferred from the orientation of the GC NTF population \citep[e.g.,][]{LaRosa2004}.

The fact that the relative angle between the FIREPLACE magnetic field and the NTFs is orthogonal indicates that the -80\degree\ component observed in \citetalias{Pare2024} is not spatially associated with the NTFs. Instead, it could be revealing a vertical field that pervades the GC. If there is a pervasive vertical field it would help discriminate between theories regarding how the seemingly distinct vertical and horizontal magnetic field systems in the GC connect. This again supports the possibility that an initially vertical (or poloidal) field could have been sheared into a horizontal configuration in the denser molecular regions of the CMZ \citep{Morris1996b}. Alternatively, the vertical field configuration could be generated from winds or outflows from the parallel field \citep{Su2010,Heywood2019}. The HRO/PRS results related to the NTFs in this work support the theory that there is a pervasive vertical field that becomes sheared in the CMZ molecular clouds.

\subsection{An Upper Limit on the Magnetic Field Strength of a Pervasive Vertical Field}
\begin{figure*}
    \centering
    \includegraphics[width=1.0\textwidth]{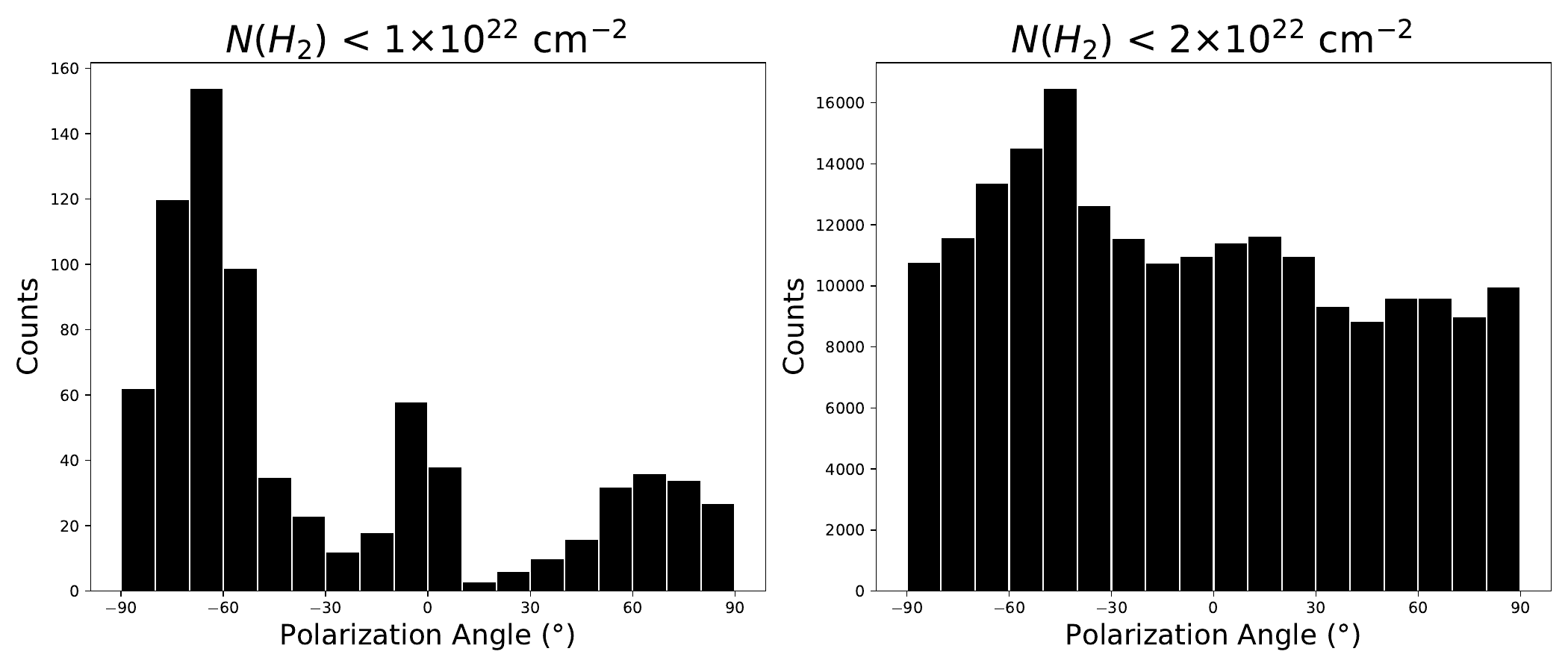}
    \caption{Histograms showing the distribution of FIREPLACE polarization angles corresponding to low column densities in the CMZ, with column density cuts of (left panel) $N(H_2) < 1\times10^{22}$ cm$^{-2}$ and (right panel) $N(H_2) < 2\times10^{22}$ cm$^{-2}$. In these histograms an angle of 0\degree\ indicates a magnetic field oriented parallel to the Galactic plane and an angle of 90\degree\ indicates a magnetic field oriented perpendicular to the Galactic plane.}
    \label{fig:B_lownh2}
\end{figure*}
The density-independent negative $V$ observed for the TLP cloud complex shown in Figure \ref{fig:multi2} is in sharp contrast with the generally positive $V$ observed for the other CMZ clouds. The TLP is also the lowest column density molecular cloud studied in this work (Table \ref{tab:cloud_param}). The alignment trend observed for the TLP could therefore indicate that lower column density regions of the CMZ are more magnetically dominated and subject to a more vertical field like the one seemingly traced by the NTFs. The TLP may represent an early stage of the shearing process where the field visibly transitions from perpendicular to parallel with respect to the cloud structure.

Building on this possibility, if we assume there is a pervasive vertical field in the CMZ that then becomes sheared into a  horizontal configuration in the higher column density regions of the CMZ, we can use our results from this work to estimate the upper limit of the magnetic field strength for this vertical field:
\begin{equation}
    B \leq \sqrt{4\pi\rho}v, \label{eq:B1}
\end{equation}
where $B$ is the strength of the magnetic field (G) in the plane of the sky, $\rho$ is the $H_2$ mass density (g cm$^{-3}$), and $v$ a characteristic orbital velocity of the CMZ molecular clouds (cm s$^{-1}$). This equation is derived from the relation $(1/2)\rho{}v^2 = (B^2/8\pi)$ that equates magnetic and kinetic energy densities.

To perform this magnetic field strength estimate, we must therefore first estimate the $H_2$ mass density. To do so, we estimate the column density level at which the magnetic field inferred from FIREPLACE transitions to a largely vertical field. We do so by inspecting the low column density regimes of the CMZ and creating histograms of the polarization angle orientation with respect to the Galactic plane. We show two example histograms in Figure \ref{fig:B_lownh2}, where the left panel shows the histogram of the FIREPLACE polarization angles coincident with $N(H_2) < 1\times10^{22}$ cm$^{-2}$. The right panel shows the same, but for all polarization angles coincident with $N(H_2) < 2\times10^{22}$ cm$^{-2}$. In this figure a polarization angle of 0\degree\ indicates a magnetic field that is oriented parallel to the Galactic plane, whereas a polarization angle of 90\degree\ indicates a magnetic field that is oriented perpendicular to the Galactic plane. 

We observe that the polarization angle distribution corresponding to the $N(H_2) < 1\times10^{22}$ cm$^{-2}$ cut shown in the left panel of Figure \ref{fig:B_lownh2} is non-uniformly distributed. This non-uniform distribution begins to reveal the vertical magnetic field seemingly traced by the NTFs, as can be seen by the peak at an angle of -70\degree. The distribution of polarization angles associated with the $N(H_2) < 2\times10^{22}$ cm$^{-2}$ cut in the right panel of Figure \ref{fig:B_lownh2} is conversely more uniformly distributed across the -90 -- 90\degree angle range. This more uniform distribution is also observed for polarization angle distributions drawn from higher column density thresholds. We therefore use $N(H_2) = 1\times10^{22}$ cm$^{-2}$ as the highest column density at which the FIREPLACE magnetic field traces the vertical field component in the GC. The true column density at which this transition occurs is likely lower than this estimate, but we use this threshold value for this calculation since we lack significant statistics for lower $N(H_2)$ cuts in this data set.

We then convert from this threshold column density to mass density: $\rho = m_{H2}\times\frac{N(H_2)}{\Delta'}$, where $m_{H_2}$ is the mass of H$_2$ (3.34$\times$10$^{-24}$ g) and $\Delta'$ is a representative thickness of the CMZ molecular clouds (in units of cm). We estimate a characteristic thickness by determining the average widths of the CMZ clouds studied in this paper in the plane-of-the-sky and assuming the clouds are cylindrically symmetric. We find that the average thickness from these clouds in the plane-of-the-sky is $\sim$50\arcsec\ ($\sim$2.0 pc). Converting to cm we get $\Delta'\sim6.3\times10^{18}$ cm. For our velocity estimate $v$ we take a representative orbital velocity of 150 km s$^{-1}$ ($1.5\times10^7$ cm s$^{-1}$) accounting for CMZ orbital velocities ranging from 120 -- 200 \kms\ \citep[e.g.,][]{Kruijssen2015}. 

We can then use our estimated values and re-write Equation \ref{eq:B1} to estimate the magnetic field strength:
\begin{equation}
    B \leq 2\sqrt{\pi{}m_{H2}}\times\left(\frac{N(H_2)}{N(H_2)_0}\right)^{1/2}\left(\frac{v}{v_0}\right) \left(\frac{\Delta'}{\Delta'_0}\right)^{-1/2}. \label{eq:B2}
\end{equation}
This form of Equation \ref{eq:B1} allows for re-scaling of the estimated $N(H_2)$, $\Delta'$, and $v$ parameters to determine magnetic field strengths. Reproducing this equation with the estimated parameters discussed above:
\begin{multline}
    B \leq (4\,\mathrm{mG})\left(\frac{N(H_2)}{1\times10^{22}\,\mathrm{cm}^{-2}}\right)^{1/2} \\ \left(\frac{v}{1.5\times10^7\,\mathrm{cm}\,\mathrm{s}^{-1}}\right)\left(\frac{\Delta'}{6.3\times10^{18}\,\mathrm{cm}}\right)^{-1/2}. \label{eq:B3}
\end{multline}
Using these values we obtain an upper limit of $B \leq 4$ mG for this pervasive vertical field.

Previous derivations of magnetic field strengths from both observations and theory have derived strengths ranging from 100s of $\mu$G to 10s of mG \citep[e.g.,][]{Pillai2015,Mangilli2019,Guerra2023,Lu2024}. Our estimate on the vertical magnetic field strength is well within this range of previously derived magnetic field strengths. We therefore cannot rule out the existence of a pervasive $\sim$mG field that exists throughout the GC, though a significantly lower field strength is also possible.

This conclusion yields important insight into questions relating to the distribution of NTFs throughout the GC: why do the NTFs appear in only certain regions of the GC if there is a pervasive field? One possible answer to this question is that the NTFs coincide with regions of enhanced magnetic field strength relative to their surroundings. Representative mechanisms for this behavior include plasma compression, magnetic field line stretching, colliding and shearing flows, or a plasma stream that encounters an obstacle \citep{Drake2013,Guo2020,Yusef-Zadeh2022spac}. These regions could generate synchrotron cooling leading to the illumination of the NTFs \citep[e.g.,][]{Yusef-Zadeh2019}.

Alternatively, the NTFs could potentially form anywhere in the GC where certain other conditions (such as proximity to a source of cosmic rays) are met. The NTFs are generally estimated to have local magnetic field strengths of 100s of $\mu$G using equipartition estimates \citep[e.g.,][]{Gray1995,Lang1999a,Lang1999b,Pare2022}. These are likely lower level estimates for the magnetic field strength, and the true field strength local to the NTFs could be much higher, even mG strength fields \citep{Morris2006,Pare2022}. The illumination of an NTF could therefore potentially occur anywhere in the GC where the necessary conditions to generate CRs arise, such as from magnetic field reconnection at the surfaces of molecular clouds or pulsar wind nebulae \citep[e.g.,][]{Serabyn1994,Thomas2020}. The mG magnetic field strength estimate found here, in conjunction with the NTF equipartition field strength estimates, cannot rule out the presence of a strong, pervasive magnetic field in the GC. It is therefore possible that the NTFs are not local enhancements of a weaker background field.

\section{CONCLUSIONS} \label{sec:conc}
In this work we have analyzed the relative angle between the FIREPLACE magnetic field and the column density structure of the CMZ. We have also compared the direction of this magnetic field with the orientation of the NTF population within the CMZ. We then studied the relative alignment for individual molecular clouds within the CMZ and compared the alignments observed for the clouds with the larger CMZ-wide results. We summarize the major conclusions from our work here:

\begin{itemize}
    \item We observe an increasingly aligned magnetic field orientation with column density structure at higher column densities for the full CMZ. This is an opposite trend as a function of column density to what is observed in Galactic Disk star forming regions \citep{PlanckXXXV,Soler2019}. The increasingly parallel magnetic field is a strong indication of significant shear occurring within CMZ molecular clouds. 
    \item We observe a range of alignment trends when studying individual CMZ clouds (the Ring Cloud, Sgr B2, Cloud e/f, the Bricklets, the Brick, the TLP cloud complex, the 20 \kms\ cloud, and Sgr C). The generally positive $V$ observed for the CMZ clouds could indicate the importance of shear resulting from the motion of the clouds through the gravitational potential of the GC. The orientation of the magnetic field is slightly dependent on the column density of the clouds, where lower density regions, like the TLP, exhibit $V<0$ which could indicate that such regions are transitioning to a gravitationally sheared phase.
    \item We see a correlation between the magnetic field alignment and the peak column densities of the individual CMZ clouds as shown in Table \ref{tab:cloud_param}. We also observe that the CMZ molecular structures with negative $V$ are likely experiencing shocks. This trend could indicate regions in the CMZ where shear caused by effects like the CMZ cloud orbital motion enable the gravitational potential to overcome the turbulence and magnetic field, even in the extreme conditions of the CMZ.
    \item We see no significant alignment of the FIREPLACE magnetic field with the GC NTF orientation. The generally perpendicular orientation indicates that the vertically aligned magnetic field component observed in \citetalias{Pare2024} does not preferentially coincide with the NTFs. This supports the possibility that there is a pervasive vertical magnetic field that becomes sheared in the high density CMZ molecular clouds.
    \item We estimate the magnetic field strength in the lower density regions of the CMZ and find a characteristic field strength of $\leq$4 mG. This field strength limit does not rule out the conclusion that strong magnetic fields are present throughout the CMZ. The NTFs may therefore not be local enhancements of a weaker background field.
\end{itemize}

\begin{acknowledgements}
    This work is based on observations made with the NASA/DLR Stratospheric Observatory for Infrared Astronomy (SOFIA). SOFIA was jointly operated by the Universities Space Research Association, Inc. (USRA), under NASA contract NNA17BF53C, and the Deutsches SOFIA Institut (DSI) under DLR contract 50 OK 2002 to the University of Stuttgart. Financial support for this work was provided by NASA through award \#09-0054 issued by USRA to Villanova University. Q. Zhang acknowledges the support from the National Science Foundation under Award No. 2206512.

\end{acknowledgements}

\facility{
    SOFIA, 
    MeerKAT, 
    Herschel
    }

\software{
    Astropy \citep{Astropy2022,Astropy2018,Astropy2013},
    Matplotlib \citep{Hunter2007}, 
    Numpy \citep{Harris2020},
    Magnetar \citep{Soler2013}
    }

\bibliography{FIREPLACE_IV}{}
\bibliographystyle{aasjournal}

\end{document}